\documentclass[twocolumn,twocolappendix,floatfix,a4paper]{aastex701}

\usepackage{times}
\usepackage{amsmath}
\usepackage[varg]{txfonts}
\usepackage{bm}

\usepackage[utf8]{inputenc} 
\usepackage[T1]{fontenc}
\usepackage{natbib}
\usepackage{nameref}
\usepackage{graphicx}
\usepackage{graphics}
\usepackage[space]{grffile}
\usepackage{latexsym}
\usepackage{amsfonts,amsmath,amssymb}
\usepackage{url}
\usepackage[utf8]{inputenc}
\usepackage{fancyref}
\usepackage{hyperref}
\usepackage{xcolor}
\usepackage[normalem]{ulem}

\shorttitle{Probing Dark Matter with Strongly Lensed BBH Mergers in the Near Future}

\shortauthors{Maity \textit{et al.}}

\newcommand{\Lambdal}{\Lambda_\ell}
\newcommand{\Dl}{D_\ell}
\newcommand{\dt}{\Delta t}
\newcommand{\zl}{z_\ell}
\newcommand{\zs}{z_s}
\newcommand{\Ds}{D_s}
\newcommand{\Dls}{D_{\ell s}}
\newcommand{\mwdm}{m_{\rm wdm}}
\newcommand{\miwdm}{m^{-1}_{\rm wdm}}
\newcommand{\keV}{\,\text{keV}}
\newcommand{\GeV}{\,\text{GeV}}

\begin{document}
	
\title{Probing Dark Matter with Strongly Lensed Binary Black Hole Mergers: Prospects in the Near Future}

\author{Koustav N. Maity}
\affiliation{International Centre for Theoretical Sciences, Tata Institute of Fundamental Research, Bangalore 560089, India}
\email{koustav.narayan@icts.res.in}

\author{Souvik Jana}
\affiliation{Department of Physics, The Chinese University of Hong Kong, Shatin, NT, Hong Kong}
\affiliation{International Centre for Theoretical Sciences, Tata Institute of Fundamental Research, Bangalore 560089, India}

\email{souvikjana@cuhk.edu.hk}

\author{Ankur Barsode}
\affiliation{International Centre for Theoretical Sciences, Tata Institute of Fundamental Research, Bangalore 560089, India}
\email{ankur.barsode@icts.res.in}

\author{Tejaswi Venumadhav}
\affiliation{Department of Physics, University of California at Santa Barbara, Santa Barbara, CA 93106, USA}
\affiliation{International Centre for Theoretical Sciences, Tata Institute of Fundamental Research, Bangalore 560089, India}
\email{teja@ucsb.edu}

\author{Parameswaran Ajith}
\affiliation{International Centre for Theoretical Sciences, Tata Institute of Fundamental Research, Bangalore 560089, India}
\email{ajith@icts.res.in}

\begin{abstract}
Gravitational-wave~(GW) transients, strongly lensed by intervening galaxies and clusters, are expected to constitute a small fraction ($\sim 0.1$--$0.5\%$) of the events detectable by ground-based detectors. A strongly lensed binary black hole (BBH) merger will produce multiple copies of the GW signal arriving at different times at the detector. The abundance and time-delay distribution of these lensed events are sensitive to the mass, density profile, and redshift distribution of the lens population, and in turn, to the underlying nature of dark matter. \citet{Jana:2024dhc} recently proposed a method to constrain the masses of warm dark matter (WDM) particles making use of the strongly lensed events detectable by next-generation GW detectors. In this work, we investigate the prospects of constraining the mass of the WDM particle using upcoming observations of the upgraded LIGO-Virgo-KAGRA network. This requires a careful modeling of the detector selection effects. The upcoming fifth observing run (O5) is expected to yield only a modest bound, since only a few lensed pairs are expected at that early stage. By the sixth observing run (O6), we forecast a bound on the WDM particle mass, $\miwdm \lesssim 0.1$--$0.2\keV^{-1}$ ($\mwdm \gtrsim 5$--$10\keV$), which is comparable to the tightest existing astrophysical constraints. Runs beyond O6 will tighten the constraint by roughly an order of magnitude. Strongly lensed GWs therefore offer a complementary, competitive, and independent probe of dark matter.
\end{abstract}

\section{Introduction}
\label{sec:introduction}

The standard cosmological model, based on a cosmological constant $\Lambda$ and cold dark matter (CDM), has successfully withstood a wide range of observational tests. These include measurements of the cosmic microwave background (CMB)~\citep{planck18_A&A}, primordial light-element abundances~\citep{Fields:2019pfx}, baryon acoustic oscillations~\citep{desi2024_dr1_bao, desi2025_dr2_bao}, large-scale structure~\citep{BOSS:2016wmc}, weak gravitational lensing~\citep{KiDS_1000, des2026}, and Type Ia supernovae~\citep{Brout_2022}. Nevertheless, several of its central ingredients---including inflation, the cosmological constant (or, more generally, dark energy), and the nature of dark matter (DM)---remain poorly understood~\citep{efstathiou2024, Di_Valentino_2025}.

Neither direct-detection experiments~\citep{Undagoitia_2015, Schumann_2019} nor indirect astrophysical searches~\citep{Bertone_2005, Gaskins_2016} have found conclusive evidence for CDM particles. Motivated by a variety of theoretical and observational considerations, a broad range of DM candidates have been proposed, with masses extending from ultralight bosons to massive primordial black holes~\citep{Bertone_2005, Arbey_2021}. A simple alternative to CDM is warm dark matter (WDM), which may be produced in the early Universe as a \textit{thermal relic}. A thermal relic is a particle species that remains in chemical equilibrium with the primordial plasma while the temperature exceeds its rest-mass energy and decouples when its interaction rate falls below the Hubble expansion rate~\citep{Bertone_2005}. Sterile neutrinos~\citep{Dodelson_1994, Shi_Fuller_1999, Boyarsky_2018} and gravitinos~\citep{Pagels_Primack_1982, gravitino_wdm_1997} provide well-motivated WDM candidates.

WDM particles decouple while still relativistic, deep in the radiation-dominated era~\citep{Bode_2001}. As the Universe expands, their momenta redshift and they eventually become non-relativistic, while retaining a residual velocity dispersion---hence the designation ``warm.'' This velocity dispersion defines a free-streaming scale. Below this scale, the thermal motion of the particles allows them to escape from growing overdensities faster than gravity can confine them, thereby inhibiting the growth of density perturbations. This is the collisionless analog of the Jeans criterion, with velocity dispersion playing the role of baryonic pressure~\citep{Hogan_2000, Viel_2005}. Free streaming therefore introduces a physical cutoff in the matter power spectrum and suppresses the formation of low-mass haloes~\citep{Schneider_2012_wdm, Schneider_2013_wdm}. The formation of smaller haloes is also delayed relative to CDM. Because these haloes collapse when the mean cosmic density is lower, they generally acquire lower characteristic densities and concentrations, and hence shallower central gravitational potentials at fixed halo mass~\citep{Alam_2001, Ludlow:2016ifl}.

Electromagnetic observations across a broad range of redshifts constrain the thermal-relic WDM particle mass. At high redshift, joint CMB and quasar-absorption measurements and the JWST galaxy UV luminosity function yield $\mwdm>2.8\keV$ and $\mwdm>3.2\keV$, respectively, although the latter depends on the uncertain efficiency of early star formation~\citep{Chatterjee_2024,Liu_2024}. The Lyman-$\alpha$ forest gives $\mwdm>5.7\keV$, subject to uncertainties in the thermal history of the intergalactic medium~\citep{Irsic2024}. At lower redshifts, Local Group satellite populations and stellar streams yield bounds of $\mwdm>6.2\keV$, with uncertainties arising from baryonic physics and the galaxy-halo connection~\citep{liu2_2026,Banik_2021}. Strong-lensing flux ratios provide a complementary, purely gravitational probe, with recent JWST observations yielding $\mwdm>9.6$--$10.2\keV$, depending on the adopted priors~\citep{gilman2026jwstlensedquasardark}. Joint analyses combining lensing, satellite populations, and the Lyman-$\alpha$ forest obtain bounds in the range $\mwdm\gtrsim6$--$10\keV$~\citep{Nadler_2021,Enzi_2021}.

Complementing these electromagnetic probes, gravitational-wave (GW) observations offer new ways to investigate DM. Over the past decade, LIGO~\citep{aasi2015advanced} and Virgo~\citep{acernese2014advanced} have observed more than four hundred GW signals from merging binaries containing black holes and neutron stars~\citep{abbott2019gwtc, abbott2021gwtc, abbott2024gwtc, ligo2023gwtc, abac2025gwtc, abac2026gwtc, venumadhav2020new, zackay2021detecting, olsen2022new, mehta2025new, wadekar2023new, nitz20191, nitz20202, nitz20213, nitz20234, koloniari2025new}. Planned sensitivity upgrades to the existing detectors~\citep{H1L1V1-psd-O3O4O5, HLA-psd-O6, HLA-voyager}, the addition of LIGO-India to the global network~\citep{LIGO-M1100296-v2, Saleem_2022}, and proposed next-generation (XG) observatories such as Cosmic Explorer and the Einstein Telescope~\citep{Reitze_2019, evans2021horizonstudycosmicexplorer, maggiore2020science, hild2011sensitivity} will increase the observable volume by orders of magnitude and could enable the detection of millions of GW events~\citep{Kalogera:2021bya, Chen:2024gdn}.

A small fraction ($\sim 0.1$--$0.5\%$) of these signals are expected to be strongly lensed by intervening galaxies and clusters, producing multiple images of the same signal that arrive at different times.\footnote{For ground-based detectors, the GW wavelength is much smaller than the characteristic scale of the relevant lenses, $\lambda_{\rm GW} \ll 2GM_{\rm lens}/c^{2}$, and diffraction effects are therefore negligible. In this \textit{geometric-optics} regime, lensing does not alter the intrinsic shape of the GW signal.}
Given the expected lensing rates, the first detection of a strongly lensed GW may occur within the next few years~\citep{barsode2025lensing}.\footnote{No evidence for strong lensing has been found in the current GW data~\citep{hannuksela2019search, LIGOScientific:2021izm, abbott2023search, ligo_scientific_collaboration_and_virgo_2024_10841987, 1992schneider, li2023targeted, mcisaac2020search, dai2020search, janquart2023follow, abac2025gwtclens, barsode2026search}.} As detector sensitivities improve, such detections should become increasingly common and may soon provide a statistically informative population of lensed events.

Because GWs interact only gravitationally with intervening structures, their strong-lensing observables directly probe the total mass distribution of the lens. In particular, the time delay ($\dt$) between lensed images encodes information about the lens mass, with lower-mass lenses generally producing shorter delays. The WDM-induced suppression of low-mass haloes should therefore deplete the short-$\dt$ tail of the observed time-delay distribution relative to the CDM prediction. \citet{Jana:2024dhc} proposed using the distribution of lensing time delays observed by XG detectors to constrain the WDM particle mass. Their analysis assumed that the detectors would be sensitive to essentially all binary black hole (BBH) mergers and that the unlensed BBH population would be accurately characterized using the much larger sample of unlensed events. Under these assumptions, they showed that strongly lensed GWs could produce substantially tighter constraints on WDM than existing observational probes.

In this work, we investigate the potential of upcoming observing runs of current-generation detectors to constrain the WDM particle mass. Unlike XG detectors, current detectors have a limited observing horizon, making a careful treatment of detector selection effects essential. We model the lensing selection function for observing scenarios ranging from the fourth LVK observing run (O4) to several future detector networks. For two astrophysically motivated BBH merger-rate distributions, we calculate the detectable lensing fraction and the lensing time-delay distribution as functions of the WDM particle mass, assuming a fiducial cosmology. Taking DM to be truly cold, we then forecast the lower bound on $\mwdm$ achievable by the end of each observing run. O5 is expected to provide only a modest constraint because it is likely to contain few lensed pairs. By O6, however, we forecast a bound of $\miwdm \lesssim 0.1\keV^{-1}$, comparable to the tightest existing astrophysical constraints. Detector networks operating beyond O6 could improve this constraint by approximately an order of magnitude.

The remainder of the paper is organized as follows. Section~\ref{sec:methods} summarizes our methodology, and Section~\ref{sec:results} presents the expected constraints. We discuss the implications of our results and future prospects in Section~\ref{sec:outlook_and_conclusion}. Appendix~\ref{app:proj_obs_runs} summarizes the detector configurations, locations, and sensitivities considered in this work. Unless specified otherwise, we use the \textsc{Planck18} cosmological parameters throughout the paper~\citep{Planck18}. We also use geometrized units: $G = c = 1$.

\section{Methods}
\label{sec:methods}

\subsection{The imprint of dark matter on strong-lensing observables}
\label{sec:imprint_of_dark_matter}

\begin{figure}[t]
	\centering
	\includegraphics[width=\columnwidth]{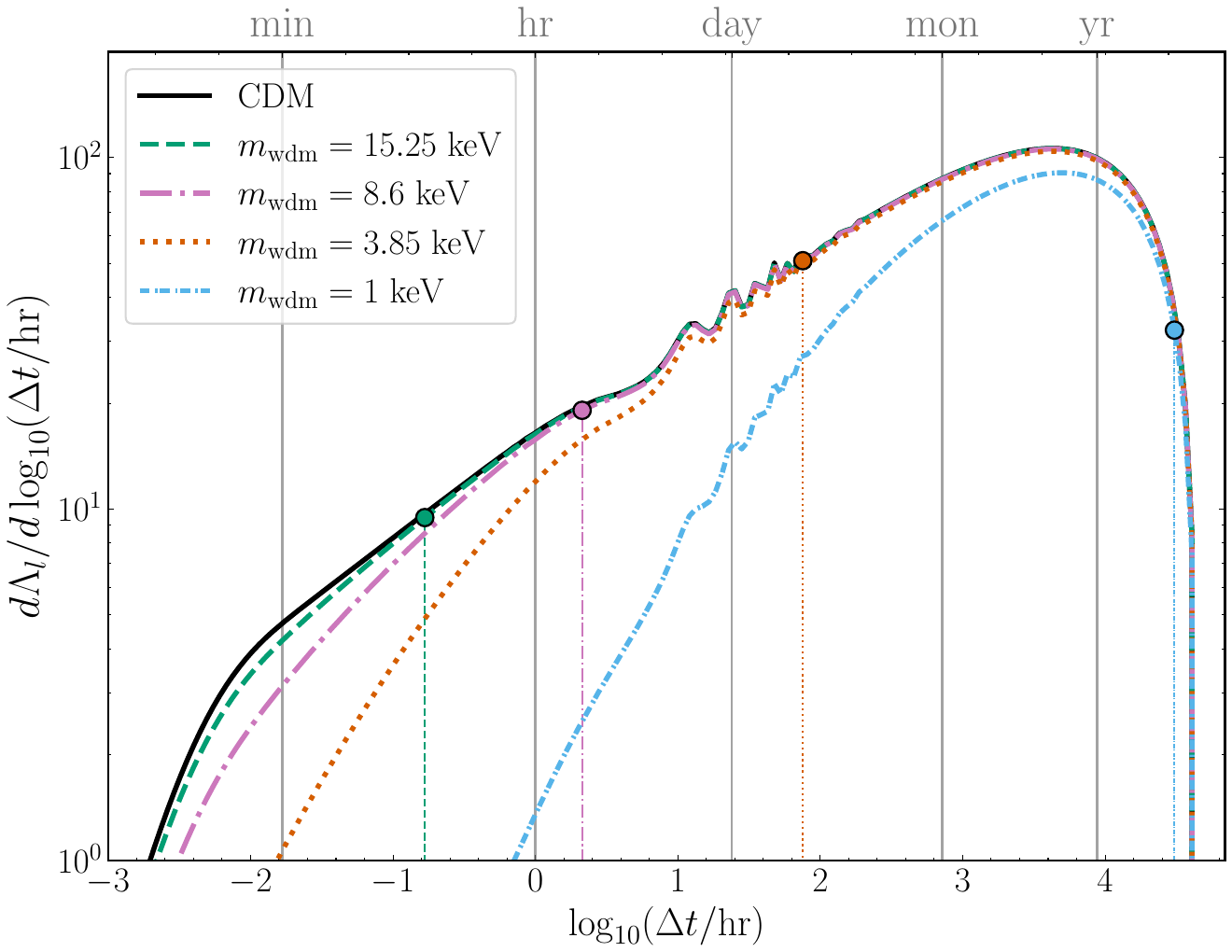}
	\caption{Expected distribution of detectable lensed events as a function of (log) $\dt$ for O6, assuming \textsc{Planck18} cosmology and the \textsc{Dominik} merger-rate model, shown for four representative WDM masses. Colored vertical lines mark, for each mass, the $\dt$ below which the WDM distribution departs from the CDM prediction.}
	\label{fig:O6_td_histograms_for_diff_mwdm}
\end{figure}

\begin{figure*}[tbh]
	\centering
	\includegraphics[width=0.85\textwidth]{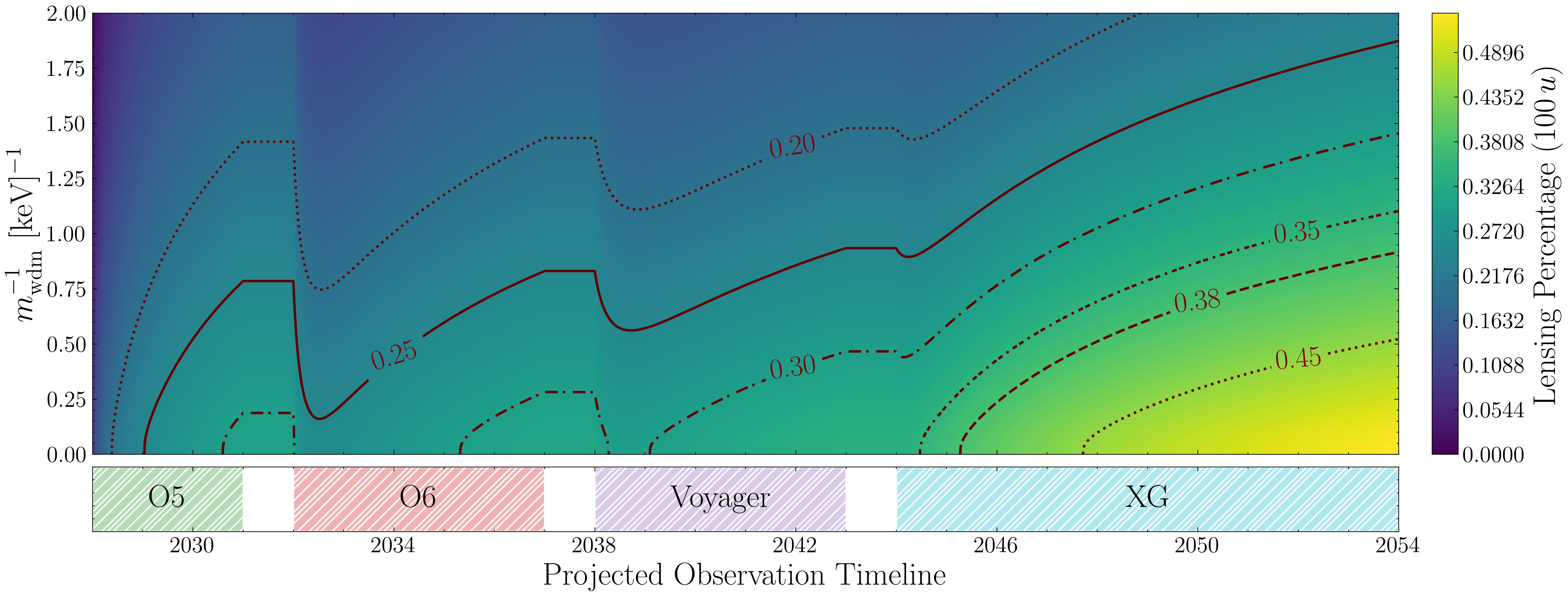}
	\caption{The cumulative percentage of strongly lensed events as a function of the projected observation timeline and WDM mass, with the timeline segmented into observing runs and detector upgrades. For the lens population we have used the \citet{Behroozi_2013} HMF, with the suppression due to WDM included via Eq.~\eqref{eq:wdm_Schnieder_modify}. For the BBH source population, we have used the \textsc{Dominik} distribution, normalized to the GWTC-3 median distribution. The timelines are illustrative and do not necessarily correspond to the anticipated schedules of future detectors.}
	\label{fig:lfraction_and_mwdm}
\end{figure*}

\begin{figure}[tbh]
	\centering
	\includegraphics[width=\columnwidth]{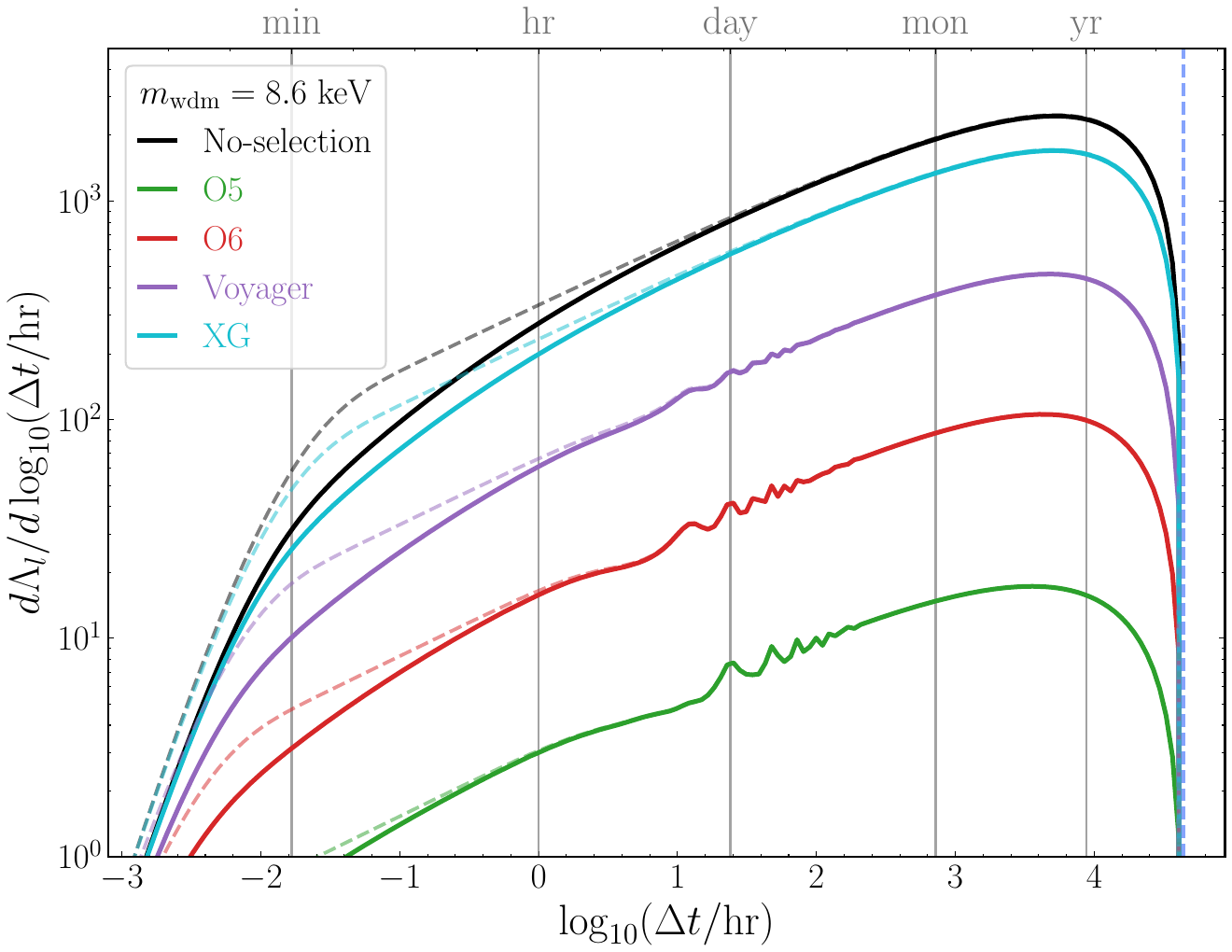}
	\caption{The time delay distribution of the detectable lensed events in the upcoming observing runs considered in this work. Thick solid lines show the expected distributions of strongly lensed pairs for $\mwdm = 8.6$ keV, while the corresponding thin dashed lines correspond to the CDM case. For both cases, as sensitivity increases (O5 $\rightarrow$ XG), a larger number of lensed pairs are detected. Also shown is the intrinsic distribution where the detector selection effects are neglected.  We assume a $5$-year observation scenario for all cases, for easier comparison, and use the \textsc{Dominik} merger rate for the BBH source population.}
	\label{fig:lensed_pair_dist_all_detect_scenario_5_yrs}
\end{figure}

The key difference between a CDM universe and a WDM universe is that free streaming suppresses the halo mass function (HMF) at low masses. The HMF is the comoving number density of DM haloes per unit mass interval. Following~\citet{Schneider_2012_wdm}, we rescale the CDM HMF ${d^{2}n_{\rm cdm}/(dM_{h} dV_{c})}$ to obtain the WDM HMF ${d^{2}n_{\rm wdm}/{(dM_{h} dV_{c})}}$:
\begin{equation}\label{eq:wdm_Schnieder_modify}
	\dfrac{d^{2}n_{\rm wdm}/(dM_{h} dV_{c})}{d^{2}n_{\rm cdm}/{(dM_{h} dV_{c})}} =\left(1 + \dfrac{M_{\rm hm} (\mwdm)}{M_{h}}\right)^{-1.16},
\end{equation}  
where $M_{\rm hm}$ is the \textit{half-mode mass scale}, at which the WDM-to-CDM abundance ratio in Eq.~\eqref{eq:wdm_Schnieder_modify} roughly falls to one half~\footnote{The half-mode scale $k_{\rm hm}$ is defined as the wavenumber at which $T(k) = \left[P_{\rm wdm}(k)/P_{\rm cdm}(k)\right]^{1/2}$ drops to half its large-scale value; the corresponding mass, $M_{\rm hm} \equiv \frac{4}{3}\pi\bar\rho\,(\lambda_{\rm hm}/2)^3$ with $\lambda_{\rm hm} = 2\pi/k_{\rm hm}$.}\citep{Viel_2005, Schneider_2012_wdm, Schneider_2013_wdm}; DM haloes with mass below $M_{\rm hm}$ will be suppressed compared to CDM. 

This suppression directly affects the population of lenses. Here, we model the lenses as singular isothermal spheres (SISs), parameterized by the velocity dispersion $\sigma$. The number of lenses per unit velocity dispersion per unit redshift follows from the HMF,
\begin{align}\label{eq:hmf_prescription}
	\dfrac{d^2n}{d\sigma d\zl}(\zl, \mwdm) = \dfrac{d^2n}{dM_{h} dV_{c}}(\zl, \mwdm) ~ \dfrac{dM_{h}}{d\sigma}(\zl) \dfrac{dV_{c}}{d\zl},
\end{align}
where ${dV_{c}}/{d\zl}$ is the differential comoving volume, while ${dM_{h}}/{d\sigma}$ is the Jacobian used to convert halo mass $M_h$ to the lens velocity dispersion $\sigma$ using the prescription of~\citep{Jana_2023,Jana_2024}. Following Eq.~\eqref{eq:wdm_Schnieder_modify}, a smaller $\mwdm$ results in a suppression of the low-$\sigma$ end of the lens distribution.   

The strong lensing optical depth---the expected probability for a GW event at a redshift $z_s$ to be strongly lensed by this lens distribution---can be computed as 
\begin{equation}
\tau(z_s, \mwdm) = \int_0^{z_s}  \int_{\sigma_\mathrm{min}}^{\sigma_\mathrm{max}} ~	\dfrac{d^2n(\zl, \mwdm)}{d\sigma d\zl}  ~ \frac{\pi r_E^2(z_s, z_l, \sigma)}{4\pi D^2_\ell(z_l)} ~ d\sigma dz_\ell. 
\label{eq:lens_opt_depth}
\end{equation}
Above, 
\begin{equation}\label{eq:einstein_radius}
r_E(\zl, \zs) = 4\pi \left(\dfrac{\sigma}{c}\right)^{2}  \dfrac{D_{\ell}(\zl) D_{\ell s}(\zl, \zs)}{D_{s}(\zs)}
\end{equation} 
is the \textit{Einstein radius} of the lens: sources that lie within $r_E$ will produce multiple images. $\Ds, \Dl$ and $\Dls$ are the angular diameter distance to the source, to the lens, and between the source and lens, respectively. $\sigma_\mathrm{min}$ and $\sigma_\mathrm{max}$ are the velocity dispersions corresponding to the least massive ($10^{8} M_{\odot}$) and most massive ($10^{15} M_{\odot}$) haloes that we assume. The probability that the source is strongly lensed \textit{at least once} then follows from Poisson statistics $P_{\ell}(\zs, \mwdm) = 1 - e^{-\tau(\zs, \mwdm)}$. More details of this calculation are available in Sec~II.C.1 of~\citet{maity2026}. 

Since more massive haloes in general produce longer $\dt$, the suppression of small-$\dt$ lensed pairs relative to CDM becomes more pronounced as $\mwdm$ decreases. Figure~\ref{fig:O6_td_histograms_for_diff_mwdm} shows the expected distribution of detectable lensed events as a function of (log) $\dt$ for the O6 observing run, corresponding to four representative WDM masses. The colored vertical lines mark, for each mass, the $\dt$ below which the WDM distribution departs from the CDM one: for $\mwdm = 15.25\keV$, only the smallest haloes are erased, so only pairs with $\dt \lesssim$ minutes are suppressed. Lowering the mass to $8.6\keV$ ($3.85\keV$) pushes this threshold out to hours (days). In the extreme case of $\mwdm = 1\keV$ (which is already ruled out by current observations), the suppression extends all the way to $\dt \sim$ years. 

We can also obtain a rough estimate of this threshold, $\dt_{\rm min}(\mwdm)$, by tracing the dependence of $\mwdm$ on $\dt$: $\mwdm$ sets $M_{\rm hm}$~\citep{Viel_2005}, setting a typical halo-mass scale; the halo-mass-to-velocity-dispersion relation~\citep{Jana_2023} converts this into a threshold $\sigma$; and the SIS $\dt$, which scales as $\sigma^4$, converts the threshold on $\sigma$ to $\dt_{\rm min}(\mwdm)$. Combining these scalings, for typical lens and source redshifts expected in O6, 
\begin{equation}
	\dt_{\rm min} \sim 3.5\,\left(\dfrac{\mwdm}{\rm keV}\right)^{-4.44} \,\rm{yr}.
\end{equation}
This estimate is accurate only at the order-of-magnitude level, but it captures why the suppression threshold moves so quickly with $\mwdm$.    

\begin{figure*}
	\centering
	\includegraphics[width=\textwidth]{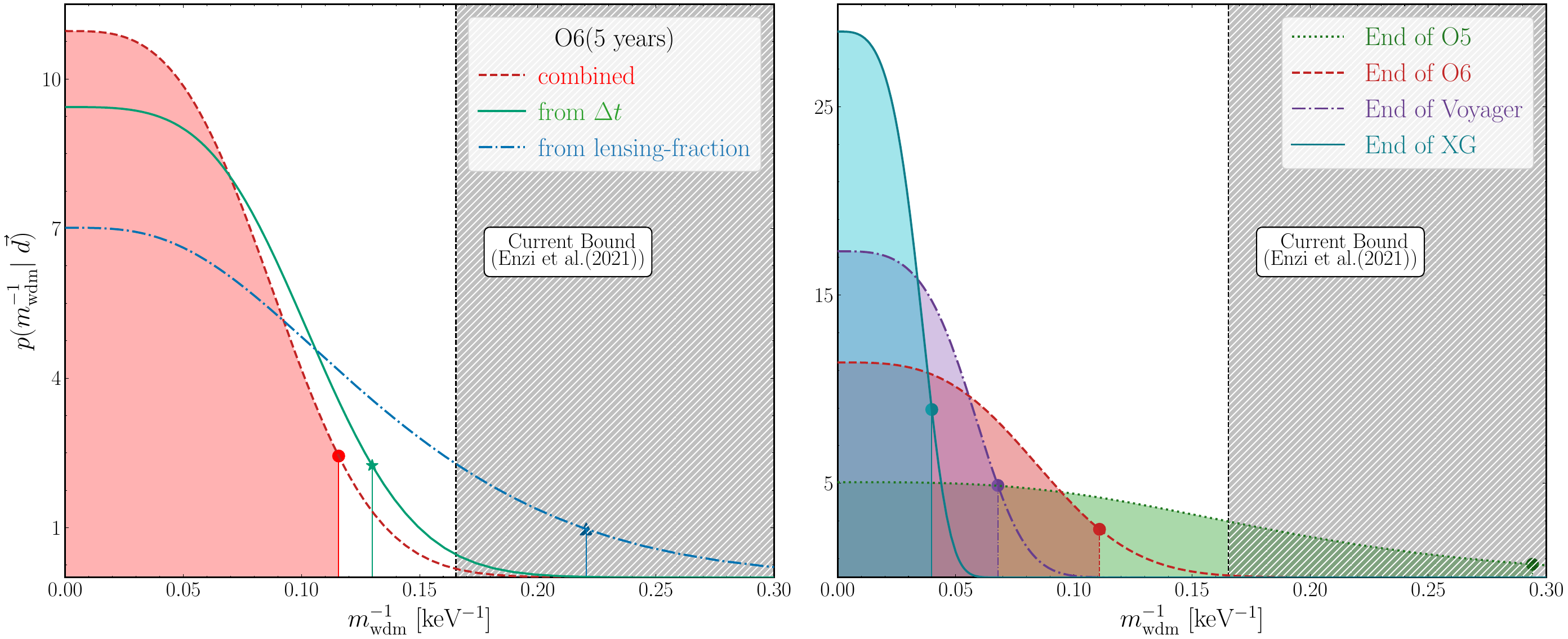}
	\caption{Posterior distributions of $\miwdm$ for the CDM scenario~($\mwdm = 1\GeV$). We used the HMF model in~\cite{Behroozi_2013} and converted to the WDM case using Eq.~\eqref{eq:wdm_Schnieder_modify} for the lens population. For the source population, we used \textsc{Dominik}~\citep{Dominik_2013}, calibrated to the GWTC-3 median rate. \textit{Left}: individual posteriors derived from $u_{\rm det}$ and $\dt$ distribution separately, along with their combined posterior, for a mock observation of $5$-year in O6. Vertical lines of the same color mark the $95\%$ upper bounds. \textit{Right}: combined posterior expected at the end of each observing run considered in this work (starting from O5), with the corresponding $95\%$ credible interval shown as shaded regions of the same color. The gray hatched area in both panels denotes the parameter space excluded by a current bound from combined analysis of different EM probes~\citep{Enzi_2021}. We note that the $95\%$ upper bound from the combined O6-only posterior (\textit{left panel}, red dashed vertical line) differs from the O6 bound shown in the right panel (red dashed vertical line), as the former is derived using only the lensed events detected during the 5-year O6 observation window, whereas the latter incorporates all lensed events accumulated through the end of the O6. Note also the difference in the vertical axis scaling between the two panels.}
	\label{fig:cdm_all_results_left_O6_right_all_det}
\end{figure*}

\subsection{Expected detection rate of lensed events and their time delay distribution}
\label{sec:strongly_lensed_expectation_td_dist}

The expected number of detectable BBH events during an observation time $T_d$ can be computed as~\citep{maity2026} 
\begin{align}\label{eq:R_det}
\Lambda =  T_d  \int_{0}^{z_{\rm max}} \dfrac{d\zs}{(1 + \zs)} ~ \dfrac{d^{2}N}{dV_{c}dt_{s}}(\zs) ~ \dfrac{dV_{c}}{d\zs}(\zs) ~ S(\zs),
\end{align}
where $d^{2}N/(dV_{c}dt_{\rm s})$ is the intrinsic BBH merger rate density per unit comoving volume in the source frame, ${dV_{c}}/{dz_s}$ is the differential comoving volume and $z_s$ is the cosmological redshift to the source. The factor $1/(1 + z_s)$ accounts for cosmological time dilation, converting from the source frame time $t_s$ to the observer frame time $t_d$. The detector selection function $S(\zs)$ encodes the probability of detecting an unlensed BBH merger at redshift $\zs$. The integral runs out to $z_{\rm max}$, the redshift beyond which we assume the intrinsic merger rate to be negligible.

To estimate the detectable number of lensed pairs, we need to compute the probability of detecting the two images, after considering the lensing optical depth and the operation times of the detectors. In computing the lensed selection function $S^L(\zs)$, we assume that if both images arrive when the detectors are operational and both exceed the $\rm{S}/\rm{N}$ threshold, they are correctly identified as a lensed pair\footnote{This is an idealized situation. In a companion paper \citep{Jana_2026_systematics}, we show that our method can still work when the identification of lensed events is imperfect. This was demonstrated in the context of GW strong lensing cosmography in \cite{Jana_2024}.}. Then, convolving all these with the strong lensing probability $P_{\ell}(\zs)$, the number of detectable strongly lensed pairs can be written as~\citep{maity2026}
\begin{align}\label{eq:exp_nl_master_eq}
	\Lambdal  = T_d  \int_{0}^{z_{\rm max}} \!\! \dfrac{d\zs}{(1 + \zs)} ~ \dfrac{d^{2}N}{dV_{c}dt_{s}} ~ \dfrac{dV_{c}}{d\zs}(\zs) ~ P_{\ell}(\zs)~S^L(\zs).
\end{align} 
Combining Eqs.~\eqref{eq:R_det} and~\eqref{eq:exp_nl_master_eq}, we can define the \textit{cumulative detectable} lensing fraction at any observation time: $u\equiv \Lambdal/\Lambda$.   

\begin{figure*}
	\centering
	\includegraphics[width=0.8\textwidth]{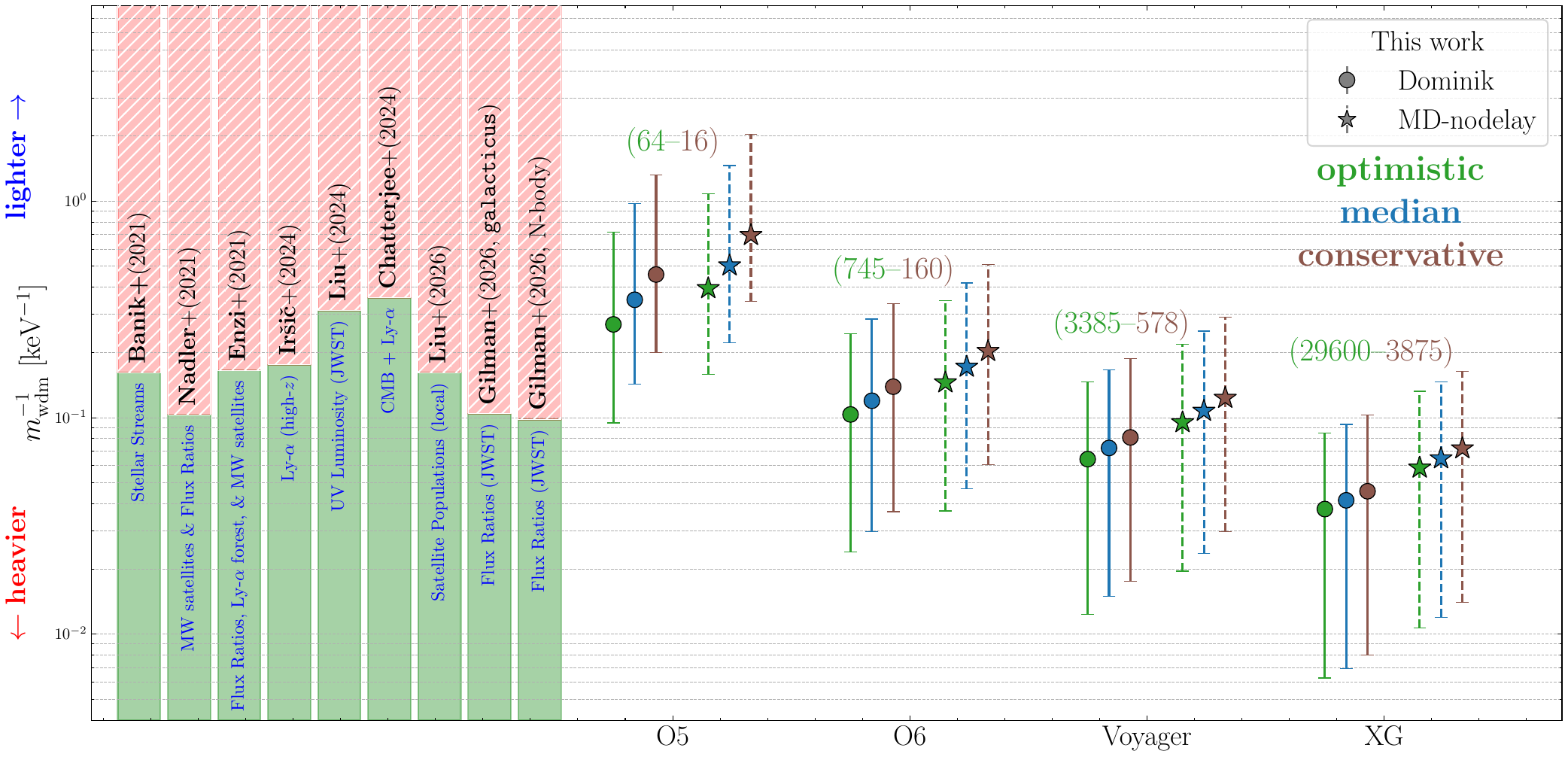}
	\caption{Forecast $95\%$ credible upper limits on $\miwdm$ for four future observing runs. The columns on the left summarize the allowed (green) and excluded (red) WDM-mass regions from EM astrophysical probes. Circles (solid error bars) assume the \textsc{Dominik} merger-rate distribution; stars (dashed error bars) assume \textsc{MD-nodelay}. Colors denote the optimistic, median, and conservative intrinsic-rate calibrations to the GWTC-3 results. Each point shows the median $95\%$ credible upper bound on $\miwdm$ across $5000$ independent realizations of the CDM universe, with error bars indicating the central 68\% range across realizations.  The numbers in bracket show the expected number of lensed events in the corresponding run, with green being the optimistic (\textsc{Dominik}, optimistic) and brown being the conservative (\textsc{MD-nodelay}, conservative) estimate.}
	\label{fig:WDM_all_constraints_summary}
\end{figure*}

The predicted strong-lensing rate depends sensitively on both the normalization and redshift evolution of the BBH merger rate~\citep{Biesiada:2014kwa,Ng:2017yiu,Oguri:2018muv,Mukherjee:2021qam}. We consider two models for its normalized redshift distribution: (1) the BBH population-synthesis model of \citet{Dominik_2013}, denoted by ``\textsc{Dominik}'', and (2) a model that follows the cosmic star-formation history of \citet{Madau_2014} with no delay between binary formation and merger, denoted by ``\textsc{MD-nodelay}''. We normalize both distributions to the local merger rate inferred from GWTC-3~\citep{ligo2023gwtc}. Our fiducial forecasts use the median GWTC-3 rate. To obtain the optimistic and conservative forecasts shown in Fig.~\ref{fig:WDM_all_constraints_summary}, we also use the upper and lower bounds of the 50\% credible interval of the rate posterior, respectively. In a companion study, we examine a broader range of merger-rate models~\citep{Jana_2026_systematics}.

Now we turn to the time-delay distribution of detectable lensed events. The lensing time delay $\Delta t$ is a function of the source redshift $z_s$, lens redshift $z_\ell$, the impact parameter $y$ as well as the velocity dispersion $\sigma$ of the lens. The time-delay distribution of detectable lensed events, ${dP^{\rm det}}/{d\dt}$ is derived by marginalizing the intrinsic distribution ${dP}/{d\dt}$ over the expected distributions of $z_s, z_\ell, y$ and $\sigma$, after appropriately weighting by the detector selection function
\begin{align}\label{eq:td_bias_master}
\dfrac{dP^{\rm det}}{d\Delta t}& \propto \int_0^{z_\mathrm{max}} \!\!\!\! d\zs \dfrac{dP_{b}}{d\zs} P_{\ell}(\zs) \int_0^1 \!\!\!\! dy \frac{dP}{dy}~ S^L(\zs, \Delta t, y) ~ \frac{dP}{d\Delta t} (\zs, y),
\end{align}
where the normalization is such that $\int_0^{T_d} d\Delta t \, {dP^{\rm det}}/{d\Delta t} = 1$. Above, $dP_{b}/d\zs$ is the redshift distribution of the merger rate density in the detector time, $P_{\ell}(\zs)$ is the strong lensing probability for a source at $z_s$, ${dP}/{dy}$ is the expected distribution of the impact parameter, $S^L(\zs, \Delta t, y)$ is the strong lensing selection function (the probability that the two images of a lensed event located at redshift $z_s$, impact parameter $y$ and time delay $\Delta t$ are detected), and ${dP} (\zs, y)/{d\Delta t}$ is the intrinsic distribution of the lensing time delay as a function of $\zs$ and $y$ after marginalizing over $\zl$ and $\sigma$. The selection function $S^L(\zs)$ in Eq.~\eqref{eq:exp_nl_master_eq} is obtained by marginalizing $S^L(\zs, \Delta t, y)$ over $\Delta t$ and $y$. Details of this calculation are given in Sec II.C of~\cite{maity2026}. Figure~\ref{fig:lfraction_and_mwdm} shows cumulative lensing percentage of strongly lensed events as a function of the projected observation timeline and WDM mass, while Fig.~\ref{fig:lensed_pair_dist_all_detect_scenario_5_yrs} shows the expected time delay distribution: $d\Lambda_\ell/d\log_{10}\Delta t = \Lambda_\ell \, dP/d\log_{10}\Delta t$.

\subsection{Bayesian inference on the nature of dark matter}
\label{sec:bayes_formalism}

We will assume that out of $N_{\rm tot}$ detected BBH signals, a subset $N_\ell$ are confidently identified as strongly lensed image \textit{pairs}, with their time-delays $\{\dt_i\}$ measured with negligible errors. If we define $N$ as the number of \textit{actual} mergers detected, then $N_{\rm tot} = N + N_\ell$. Our observable set is therefore the detected lensing fraction\footnote{The convention we adopt here is consistent with the definition of the expected lensing fraction, $u = \Lambdal/\Lambda$.} $u_{\rm det} \equiv N_\ell/N$ and the distribution of time-delay samples $\{\dt_i\}$. We calculate the posterior for the inverse of WDM particle mass $\miwdm$ using Bayes' theorem\footnote{We compute the posterior on $\miwdm$ because it has a convenient lower bound of zero, corresponding to the CDM limit ($\mwdm \gg \keV$).}:
\begin{align}\label{eq:bayes-Master}
	p(\miwdm|~u_{\rm det}, \{\dt_i\}) = \frac{\pi(\miwdm)~p(u_{\rm det}, \{\dt_i\}|~\miwdm)}{Z},
\end{align}
where $\pi(\miwdm)$ is the prior on the inverse WDM mass, assumed to be uniform between zero and a sufficiently large upper limit that does not truncate the likelihood, and $Z$ is the Bayesian evidence, which encodes assumptions such as the choice of HMF and lens model. Conditional on $\miwdm$, we factorize the likelihood as
\begin{equation}\label{eq:lhd-Master}
	p(u_{\rm det}, \{\dt_i\}|~\miwdm) = p(u_{\rm det}|~\miwdm)~p(\{\dt_i\}|~\miwdm).
\end{equation}

Following~\cite{maity2026}, we can write, 
\begin{align}\label{eq:lfrac_lhd_concurrent}
	p(u_\mathrm{det} | \miwdm) & \simeq  \int_{0}^{\infty} \!\! dN \int_{0}^{N} \!\! dN_\ell \, \delta \left(u_\mathrm{det} - \dfrac{N_\ell}{N}\right) \nonumber \\
	&\times p\left(N |\Lambda (\miwdm)\right) p\left(N_\ell | \Lambdal(\miwdm) \right), 
\end{align}      
where $p\left(N |\Lambda (\miwdm)\right)$ and $p\left(N_\ell | \Lambdal(\miwdm)\right)$ are Poisson likelihoods with $\Lambda (\miwdm)$ and $\Lambdal(\miwdm)$ calculated using Eqs.~\eqref{eq:R_det} and \eqref{eq:exp_nl_master_eq} respectively
\footnote{Note that $N = N_{\ell} + N_{u}$, where $N_{u}$ is the number
	of detected unlensed events, so $N_{\ell}$ and $N$ are not independent.
	The variables that do factorize are $N_{\ell}$ and $N_{u}$. The exact
	likelihood is
	\begin{align}
		p(u_\mathrm{det} \mid \miwdm) =
		\int_{0}^{\infty} \!\! dN_{u} \int_{0}^{\infty} \!\! dN_{\ell} \;
		\delta \left(u_\mathrm{det} - \frac{N_{\ell}}{N_{\ell} + N_{u}} \right)
		\nonumber \\
		\times\, p\big( N_{u} \mid (1 - u)\Lambda(\miwdm)\big) \,
		p\big( N_{\ell} \mid \Lambdal(\miwdm) \big) , \nonumber
	\end{align}
	where $u \equiv \Lambdal/\Lambda$ is the expected lensing fraction, not
	to be confused with its mock realization $u_\mathrm{det}$. Since $u \ll 1$,
	this differs from Eq.~\eqref{eq:lfrac_lhd_concurrent} only at order
	$u$. Neglecting the correlation inflates rather than shrinks the error
	bar, so the approximation is conservative.}.
The likelihood of observing a time-delay $\dt_i$ is given by ${dP^{\rm det}}/{d\dt}$ evaluated at $\dt_i$. Therefore,
\begin{equation}\label{eq:td_lhd_eq1}
	p\left(\lbrace \dt_i\rbrace|~ \miwdm \right) = 
	\prod_{i=1}^{N_{\ell}} \frac{dP^{\rm det}}{d\dt}\left(\dt_i~|~\miwdm \right),  
\end{equation}
as individual lensed events are statistically independent. 

Note that our choice of a prior uniform in $\mwdm^{-1}$ corresponds to $\pi(\mwdm)\propto \mwdm^{-2}$ and therefore assigns relatively little prior weight to colder DM models with large $\mwdm$. This is a conservative choice: alternative priors that are uniform or log-uniform in $\mwdm$ place greater weight on colder models and are expected to yield tighter upper bounds on $\mwdm^{-1}$. We examine this prior dependence explicitly in a companion study~\citep{Jana_2026_systematics}.

\section{Expected constraints from strong lensing}
\label{sec:results}

\begin{figure*}
	\centering
	\includegraphics[width=\textwidth]{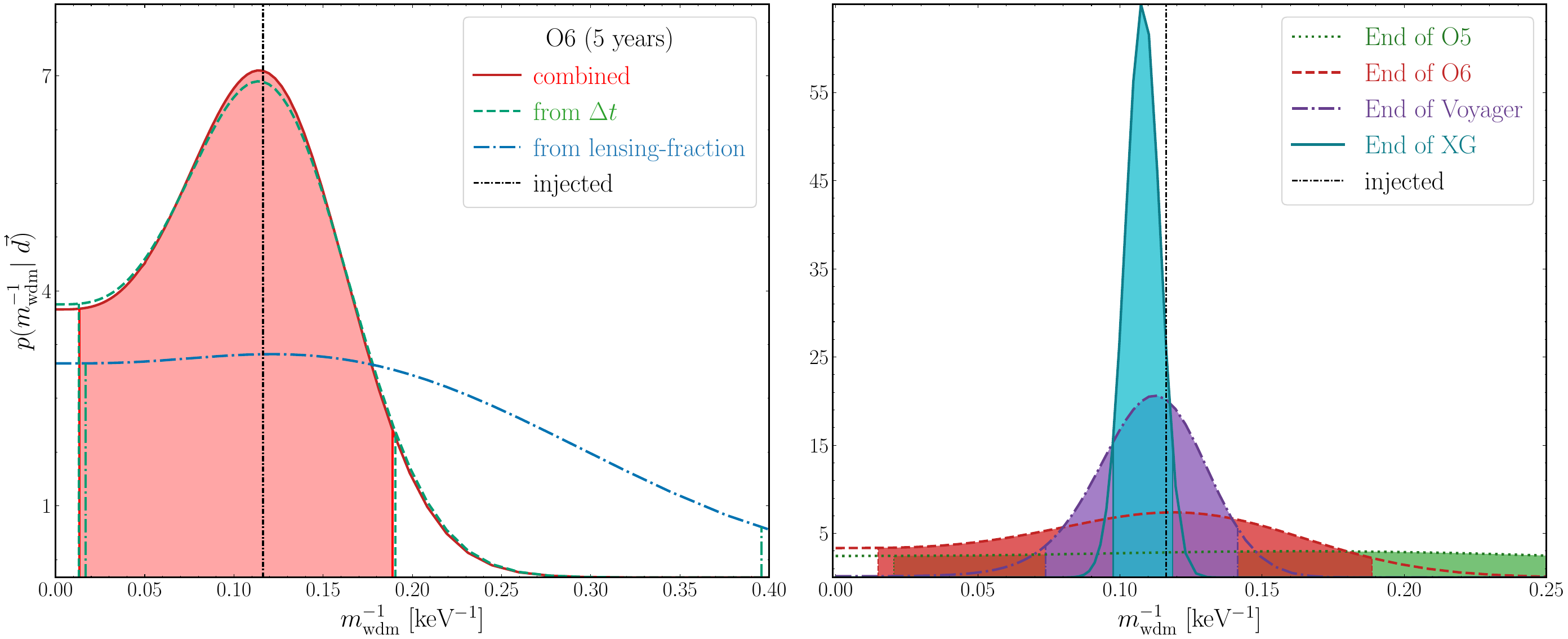}
	\caption{Inferred posterior distributions of $\miwdm$ from a simulated catalog of observations with $\mwdm = 8.6\keV$. Here, we use the \textsc{Dominik} source population model calibrated with GWTC-3 median values. Similar to Fig.~\ref{fig:cdm_all_results_left_O6_right_all_det}, left panel shows all three posteriors for a five-year mock observation in O6, with the corresponding $95\%$-credible interval shown as vertical lines. The right panel shows the combined posterior at the end of each observing run considered in this work, showing the evolution of the precision. The shaded regions are the $95\%$ credible intervals. Note also the difference in the vertical axis scaling between the two panels.}
	\label{fig:wdm_all_results_left_O6_right_all_det}
\end{figure*}

\begin{figure}[t]
	\centering
	\includegraphics[width=\columnwidth]{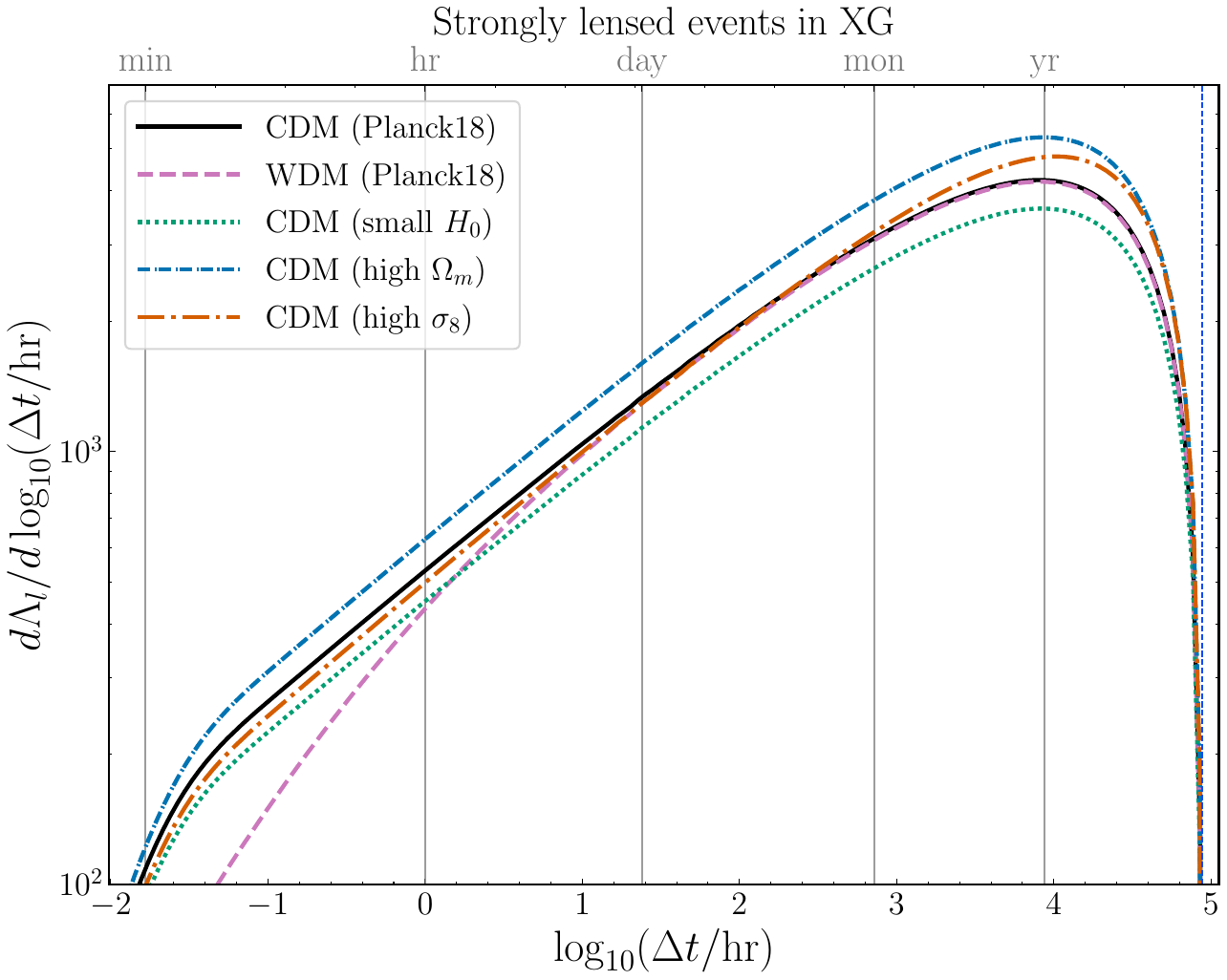}
	\caption{Expected number of detectable lensed events as a function of (log) time delay in XG detectors. The solid black curve assumes CDM in \textsc{Planck18} cosmology, while the pink dashed curve assumes WDM with same cosmological parameters. The remaining dashed curves assume CDM with one cosmological parameter changed at a time: a smaller $H_0$ (green), a higher $\Omega_{m}$ (blue), and a higher $\sigma_{8}$ (orange). Changing a cosmological parameter distorts the distribution over the whole range of $\Delta t$. Changing $\mwdm$ instead suppresses only the short $\Delta t$ end, leaving the larger $\Delta t$ region intact.}
	\label{fig:cosmo_corr_with_wdm_mass_rationale}
\end{figure}

In this section, we first present constraints on $\miwdm$ expected at the end of each observing run assuming that DM is actually cold. The smaller the upper limit, the more massive the DM particle, approaching the CDM. We follow a similar approach to that in~\cite{maity2026}, creating a mock observation dataset with $\mwdm=\!\!1\GeV$\footnote{One cannot set infinite WDM mass \textit{numerically} to mimic CDM, so we set a very high value. Within machine precision, it is equivalent to the CDM scenario.}. We sample the number of actual detectable BBH mergers, $N^{\rm tr}$, and lensed events, $N_\ell^{\rm tr}$, from their corresponding Poisson means (Eqs.~\eqref{eq:R_det} and~\eqref{eq:exp_nl_master_eq}), respectively. We then draw $\dt$ samples from the distribution ${dP^{\rm det}}/{d\dt}$ for the same $\mwdm$, ignoring the measurement error. Varying $\miwdm$, we compute the corresponding lensing fractions and create the $\dt$ distribution templates for the likelihoods in Eq.~\eqref{eq:lhd-Master}.

In the left panel of Fig.~\ref{fig:cdm_all_results_left_O6_right_all_det}, we show the individual posterior distributions on $\miwdm$ from the lensing fraction and the $\dt$ distribution, as well as their combination, in a $5$-year observation period in O6. While the constraint from the lensing fraction alone is modest, once the $\dt$ posterior is included, the combined posterior sets $\miwdm < 0.12\keV^{-1}$ ($95\%$ credible level). In the right panel, we demonstrate the evolution of this upper bound at the end of each observing run we have considered in this work (starting from O5). After O5, since the $u_{\rm det}$ is comparatively small, this method yields a rather modest constraint $\miwdm < 0.30\keV^{-1}$. As we accumulate lensed events, we get smaller and smaller upper bounds on $\miwdm$, eventually reaching $\miwdm < 0.036\keV^{-1}$ for XG detectors, comparable to the forecast of~\citet{Jana:2024dhc}. 

Single realizations can often be subject to Poisson fluctuations. So we generate $5000$ independent CDM universes, draw lensed and unlensed events from their Poisson means, sample $\dt$'s from the expected distribution for each scenario (O5, O6, Voyager, XG), and run the full inference. Figure~\ref{fig:WDM_all_constraints_summary} reports the median of the $95\%$ credible upper bound on $\miwdm$ (with error bars marking the $68\%$ credible interval) across these realizations, for each observing scenario and for the two merger-rate models and calibrations considered here. 

The green columns on the left of Fig.~\ref{fig:WDM_all_constraints_summary} summarize the allowed WDM-mass regions from the other astrophysical probes discussed in the introduction. The results from the strongly lensed population show the same trend as in Fig.~\ref{fig:cdm_all_results_left_O6_right_all_det}: after O6, the $95\%$ credible upper bound is $\lesssim 0.1$--$0.2\keV^{-1}$, depending on the source population and rate calibration, competitive with the current bounds. O5 forecasts stay slightly worse than the current bounds and, due to the large number of lensed events accumulated, the precision in the Voyager and XG eras is better by almost an order of magnitude. In a companion paper~\citep{Jana_2026_systematics}, we also marginalize over the inferred source population, accounting for the uncertainty in reconstructing it from the unlensed events. We find that this has little effect on the constraints, as long as the reconstruction is unbiased.

We can also ask the following question: If the DM is actually warm, how well can we estimate its mass? We investigate this by creating a mock observation catalog assuming $\mwdm = 8.6\keV$, and try to infer the WDM mass using the same likelihood in Eq.~\eqref{eq:lhd-Master}. In the left panel of Fig.~\ref{fig:wdm_all_results_left_O6_right_all_det}, we show the corresponding posteriors from lensing fraction, $\Delta t$ distribution along with their combined one for one realization of a 5-year observation in O6. The evolution of this precision is shown in the right panel. As we accumulate more and more lensed events~(O5 $\rightarrow$ XG), the $\Delta t$ distribution is more finely probed and the product of Poisson likelihoods becomes more and more sharp, resulting in smaller relative errors in the inference.      

\begin{figure*}
	\centering
	\includegraphics[width=\textwidth]{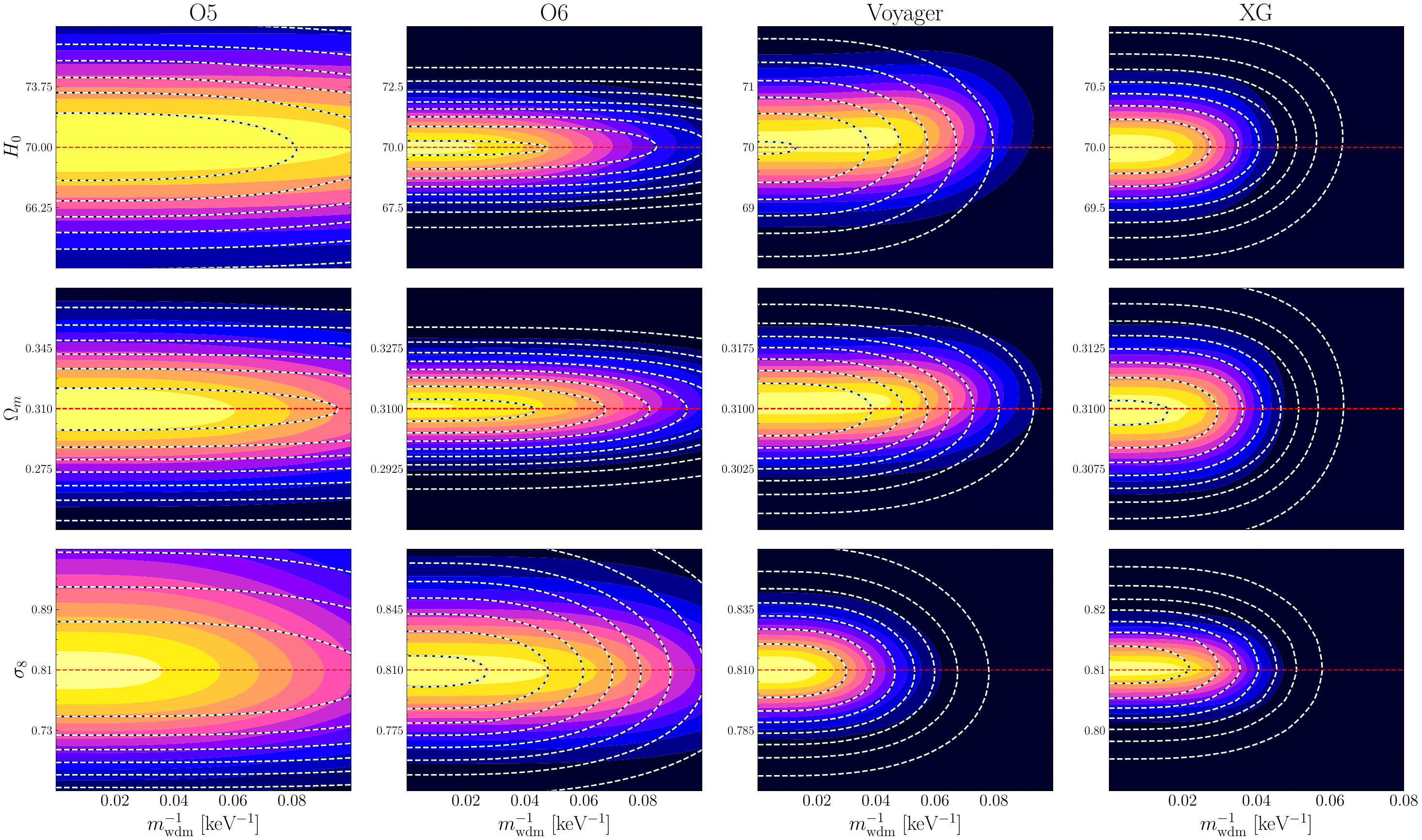}
	\caption{Two-dimensional likelihoods of $\miwdm$ and cosmological parameters, for a mock population generated assuming CDM ($\mwdm = 1$~GeV) and the fiducial cosmology. Columns correspond to the four observing scenarios (O5, O6, Voyager, and XG). In each plot, the horizontal axis corresponds to $\miwdm$ while the vertical axis corresponds to the cosmological parameters $H_0$, $\Omega_{m}$, and $\sigma_{8}$. The filled contours are for the \textsc{Dominik} source population with brighter color denoting higher likelihood values. The contour lines are for the \textsc{MD-nodelay} population model. The contours are horizontal in every panel, so $\miwdm$ has no significant degeneracy with any of the cosmological parameters. Note the different vertical limits across the columns. Both the $\miwdm$ as well as the cosmological parameters are progressively better constrained from O5 to XG, while remaining uncorrelated.}
	\label{fig:wdm_all_corr}
\end{figure*}

Several other factors can limit our ability to constrain the WDM mass. One of them is the possible degeneracy with cosmological parameters. The strong-lensing observables depend on the background cosmology, apart from the nature of DM~\citep{Jana_2023, Jana_2024, Jana:2024dhc, maity2026}. We therefore ask whether the signature of WDM on the lensing observables is degenerate with cosmological parameters $H_0$, $\Omega_{m}$, or $\sigma_{8}$. Figure~\ref{fig:cosmo_corr_with_wdm_mass_rationale} shows how the $\Delta t$ distribution is affected when the cosmological parameters and nature of DM are varied. Cosmological parameters enter the lensing time delay through the distance-redshift relation and the HMF. They distort the $\Delta t$ distribution across its entire range. However, WDM suppresses the formation of haloes below the half-mode mass, resulting in a suppression of the lower end of the $\Delta t$ distribution, leaving the higher end intact. The two effects are characteristically different, and hence are distinguishable. We now show this non-degeneracy explicitly. We generate a mock population of lensed BBH events assuming the standard values of the cosmological parameters used in this work, along with the assumption of CDM. We then compute the likelihood by simultaneously varying $\miwdm$ and the cosmological parameters. Figure~\ref{fig:wdm_all_corr} shows the combined two-dimensional likelihoods at the end of each observing scenario. The likelihoods are aligned with the horizontal axis in every panel, showing that $\miwdm$ has no significant degeneracy with $H_0$, $\Omega_{m}$, and $\sigma_{8}$. Thus, marginalizing over the cosmological parameters will not change the posterior of $\miwdm$ significantly.

\section{Summary and Future Work}
\label{sec:outlook_and_conclusion}

By probing the mass and redshift distributions of intervening lenses, strongly lensed BBH mergers can shed light on the nature of DM. Because WDM suppresses the abundance of low-mass haloes, it depletes the short-time-delay tail of the lensing time-delay distribution (Fig.~\ref{fig:O6_td_histograms_for_diff_mwdm}). Building on the proof of concept presented by \citet{Jana:2024dhc}, we forecast the constraints on the WDM particle mass achievable with the upgraded LVK network and its successors, accounting for detector selection effects and gaps in observations. We perform a Bayesian analysis of simulated populations, combining the detectable lensing fraction, $u_{\rm det}$, with the corresponding time-delay distribution. The expected upper bound improves from $\miwdm < 0.3\keV^{-1}$ in O5 to $\miwdm < 0.1$--$0.2\keV^{-1}$ in O6, and tightens by approximately an order of magnitude in the Voyager and XG eras, approaching the bounds forecast by \citet{Jana:2024dhc}. From O6 onward, the expected constraints are competitive with, and in some scenarios stronger than, those from electromagnetic probes of WDM. These forecasts are only weakly dependent on the merger-rate models and rate calibrations considered here (Fig.~\ref{fig:WDM_all_constraints_summary}). In a companion paper~\citep{Jana_2026_systematics}, we consider a broader range of merger-rate distributions.

We also show that, if DM is warm, these observations could measure its particle mass (Fig.~\ref{fig:wdm_all_results_left_O6_right_all_det}). The time-delay distribution could, in principle, constrain other DM scenarios also, including fuzzy and self-interacting DM, which modify small-scale structure in characteristically different ways~\citep{Hui_2016,MiguelRocha_sidm_2013,Tulin_2017}.

We model the lenses as SISs. Although simplified, this description is reasonably well motivated at the population level, particularly for the massive early-type galaxies that dominate the strong-lensing optical depth~\citep{Treu_Koopmans2004,Koopmans_2006,Koopmans_2009,Xu_2022,Etherington_2023}. The approximation is less reliable for low-mass, low-$\sigma$ haloes, in which the baryonic component becomes subdominant. In a companion study~\citep{Jana_2026_systematics}, we examine the effect of a central core, as predicted by some WDM models, using a non-singular isothermal-sphere lens model. That study also investigated several other potential systematics, including uncertainties in the reconstruction of the merger-rate distribution, inaccurate modeling of the HMF, and contamination of the lensed-event catalog arising from imperfect identification of lensed pairs.

Nevertheless, the lens model---and particularly the mapping from halo mass to lens parameters---should be refined substantially. A more realistic treatment could replace the SIS with a composite lens model based on the stellar-to-halo mass relation, with separate profiles for the DM and baryonic components. Subhaloes could be incorporated through a subhalo mass function, providing a smooth transition from galaxy- to cluster-scale lensing~\citep{Abe2025}. The WDM suppression of the HMF could also be generalized from the one-parameter fit adopted here [Eq.~\eqref{eq:wdm_Schnieder_modify}] to the three-parameter form proposed by \citet{Lovell2020a}, extending the analysis beyond simple thermal-WDM models. Our ongoing work aims to develop such realistic lens models, including the effects of WDM on halo density profiles. On the observational side, we use only the lensing time delays. Future analyses could incorporate complementary observables, such as image magnification ratios, to better constrain the lens population and strengthen the inference. 

\section*{Acknowledgments}
We are grateful to Alvin K. Y. Li for useful comments on the manuscript. We also thank the members of the ICTS Astrophysical Relativity group and the LVK Lensing group for useful discussions. We acknowledge support from the Department of Atomic Energy, Government of India, under projects No. RTI4019 and No. RTI4013. SJ acknowledges grants from the Research Grants Council of Hong Kong (Project No. CUHK 14304622, 14307923 and 14307724) as well as the Postdoctoral Fellowship Scheme. Numerical simulations were performed on the Alice computing cluster at the International Centre for Theoretical Sciences, Tata Institute of Fundamental Research.
This work makes use of the \texttt{HMFcalc}~\citep{murray2013hmf}, \texttt{GWFAST}~\citep{Iacovelli:2022bbs, Iacovelli:2022mbg}, \texttt{numpy}~\citep{harris2020array}, \texttt{scipy}~\citep{2020SciPy-NMeth}, \texttt{bilby}~\citep{bilby_paper}, \texttt{astropy}~\citep{astropy:2022}, \texttt{matplotlib}~\citep{Hunter:2007}, \texttt{HTcondor}~\citep{Thain:2005}, and \texttt{pyCBC}~\citep{alex_nitz_2024_10473621} software packages. 

\newpage
\appendix

\section{Projected Observing Runs and Their Sensitivities}
\label{app:proj_obs_runs}
\begin{table*}[t]
	\centering
	\renewcommand{\arraystretch}{1.1}
	\begin{tabular}{cccc}
		\hline \hline
		\begin{tabular}[c]{@{}c@{}}Detector Network\end{tabular} &
		Detector locations (sensitivity) &
		References to Noise Curves &
		\begin{tabular}[c]{@{}c@{}}Observing period (years)\end{tabular} \\ \hline
		O4    & \begin{tabular}[c]{@{}c@{}}LIGO Hanford (O4a), \\LIGO Livingston (O4a), \\ Virgo Italy (O4)\end{tabular}
		&\begin{tabular}[c]{@{}c@{}}\cite{HLV-psd-O4a} \\ \cite{HLV-psd-O4a} \\ \cite{HLV-psd-O4a}\end{tabular} & 2 \\ \hline
		IR1      & \begin{tabular}[c]{@{}c@{}}LIGO Hanford (O4), \\LIGO Livingston (O4), \\ Virgo Italy (O4)\end{tabular}
		& \begin{tabular}[c]{@{}c@{}}\cite{H1L1V1-psd-O3O4O5} \\ \cite{H1L1V1-psd-O3O4O5} \\ \cite{H1L1V1-psd-O3O4O5}\end{tabular}      & 0.5 \\ \hline
		O5      & \begin{tabular}[c]{@{}c@{}}LIGO Hanford (A+), \\LIGO Livingston (A+), \\ Virgo Italy (AdV+),  \\KAGRA Japan (KAGRA+)\end{tabular} 
		& \begin{tabular}[c]{@{}c@{}}\cite{H1L1V1-psd-O3O4O5} \\ \cite{H1L1V1-psd-O3O4O5} \\ \cite{H1L1V1-psd-O3O4O5} \\ \cite{H1L1V1-psd-O3O4O5}\end{tabular} & 3 \\ \hline
		O6      & \begin{tabular}[c]{@{}c@{}}LIGO Hanford (A$^{\sharp}$), \\LIGO Livingston (A$^{\sharp}$),\\ LIGO India (A$^{\sharp}$)\end{tabular}
		& \begin{tabular}[c]{@{}c@{}}\cite{HLA-psd-O6} \\ \cite{HLA-psd-O6} \\ \cite{HLA-psd-O6} \end{tabular} & 5 \\ \hline
		Voyager & \begin{tabular}[c]{@{}c@{}}LIGO Hanford (Voyager), \\LIGO Livingston (Voyager),\\ LIGO India (Voyager)\end{tabular}
		& \begin{tabular}[c]{@{}c@{}}\cite{HLA-voyager} \\ \cite{HLA-voyager} \\ \cite{HLA-voyager} \end{tabular} & 5 \\ \hline
		XG &
		\begin{tabular}[c]{@{}c@{}} Einstein Telescope Italy (ET-D), \\ Cosmic Explorer US (CE1 - 40 km), \\
			Cosmic Explorer US (CE2 - 20 km)\end{tabular} 
		& \begin{tabular}[c]{@{}c@{}}\cite{Borhanian_2021} \\ \cite{Reitze_2019} \\ \cite{Reitze_2019} \end{tabular} & 10 \\ 
		\hline \hline 
	\end{tabular}
	\caption{Current (O4) and proposed (IR1 and beyond) observing runs considered in this work. We assume LIGO-India will join the detector network during O6 with sensitivity comparable to the Advanced LIGO detectors at that time. For Cosmic Explorer, we model one detector with 20~km arm length and another with 40~km arm length. Columns 3 and 4 provide references for the sensitivity curves and the assumed observation periods for each observing run, respectively.}
	\label{tab:networks}
\end{table*}

Here we describe the projected observing scenarios considered in this work. After the fourth observing run~(O4), LVK detectors are undergoing upgrades~\citep{ligocurrentPlan}. An intermediate observing run (IR1) is planned to start in late 2026 and is expected to last for six months, with sensitivity comparable to O4~\citep{ligocurrentPlan}. The fifth observing run, O5, scheduled for the late 2020s, will also include KAGRA. Beyond O5, proposed upgrades such as $A^{\sharp}$ and Voyager aim to push sensitivity toward the limits of existing facility infrastructure over the coming decades; this phase is expected to also incorporate LIGO-India. Ultimately, XG detectors like Cosmic Explorer and the Einstein Telescope are expected to observe compact binaries throughout cosmic history. We adopt an optimistic projection on the observation timeline after IR1: O5 begins in 2028, followed by O6, Voyager, and XG, with commissioning breaks between each run. In Table~\ref{tab:networks} and Figure~\ref{fig:lfraction_and_mwdm}, we summarize the observation scenarios we have considered, along with their anticipated sensitivities and observation time.

\bibliography{references}

@article{Biesiada:2014kwa,
    author = "Biesiada, Marek and Ding, Xuheng and Piorkowska, Aleksandra and Zhu, Zong-Hong",
    title = "{Strong gravitational lensing of gravitational waves from double compact binaries - perspectives for the Einstein Telescope}",
    eprint = "1409.8360",
    archivePrefix = "arXiv",
    primaryClass = "astro-ph.HE",
    doi = "10.1088/1475-7516/2014/10/080",
    journal = "JCAP",
    volume = "10",
    pages = "080",
    year = "2014"
}

@article{Ng:2017yiu,
    author = "Ng, Ken K. Y. and Wong, Kaze W. K. and Broadhurst, Tom and Li, Tjonnie G. F.",
    title = "{Precise LIGO Lensing Rate Predictions for Binary Black Holes}",
    eprint = "1703.06319",
    archivePrefix = "arXiv",
    primaryClass = "astro-ph.CO",
    doi = "10.1103/PhysRevD.97.023012",
    journal = "Phys. Rev. D",
    volume = "97",
    number = "2",
    pages = "023012",
    year = "2018"
}

@article{Oguri:2018muv,
    author = "Oguri, Masamune",
    title = "{Effect of gravitational lensing on the distribution of gravitational waves from distant binary black hole mergers}",
    eprint = "1807.02584",
    archivePrefix = "arXiv",
    primaryClass = "astro-ph.CO",
    doi = "10.1093/mnras/sty2145",
    journal = "Mon. Not. Roy. Astron. Soc.",
    volume = "480",
    number = "3",
    pages = "3842--3855",
    year = "2018"
}

@article{Mukherjee:2021qam,
    author = "Mukherjee, Suvodip and Broadhurst, Tom and Diego, Jose M. and Silk, Joseph and Smoot, George F.",
    title = "{Impact of astrophysical binary coalescence time-scales on the rate of lensed gravitational wave events}",
    eprint = "2106.00392",
    archivePrefix = "arXiv",
    primaryClass = "gr-qc",
    doi = "10.1093/mnras/stab1980",
    journal = "Mon. Not. Roy. Astron. Soc.",
    volume = "506",
    number = "3",
    pages = "3751--3759",
    year = "2021"
}

@article{Planck18,
  title={Planck 2018 results-VI. Cosmological parameters},
  author={Aghanim, Nabila and Akrami, Yashar and Ashdown, Mark and Aumont, J and Baccigalupi, C and Ballardini, M and Banday, AJ and Barreiro, RB and Bartolo, N and Basak, S and others},
  journal={Astronomy \& Astrophysics},
  volume={641},
  pages={A6},
  year={2020},
  publisher={EDP sciences}
}

@article{BOSS:2016wmc,
	author = "Alam, Shadab and others",
	collaboration = "BOSS",
	title = "{The clustering of galaxies in the completed SDSS-III Baryon Oscillation Spectroscopic Survey: cosmological analysis of the DR12 galaxy sample}",
	eprint = "1607.03155",
	archivePrefix = "arXiv",
	primaryClass = "astro-ph.CO",
	doi = "10.1093/mnras/stx721",
	journal = "Mon. Not. Roy. Astron. Soc.",
	volume = "470",
	number = "3",
	pages = "2617--2652",
	year = "2017"
}

@article{Fields:2019pfx,
    author = "Fields, Brian D. and Olive, Keith A. and Yeh, Tsung-Han and Young, Charles",
    title = "{Big-Bang Nucleosynthesis after Planck}",
    eprint = "1912.01132",
    archivePrefix = "arXiv",
    primaryClass = "astro-ph.CO",
    reportNumber = "UMN--TH--3902/19, FTPI--MINN--19/25",
    doi = "10.1088/1475-7516/2020/03/010",
    journal = "JCAP",
    volume = "03",
    pages = "010",
    year = "2020",
    note = "[Erratum: JCAP 11, E02 (2020)]"
}

@article{planck18_A&A,
	title={Planck 2018 results-I. Overview and the cosmological legacy of Planck},
	author={Aghanim, Nabila and Akrami, Yashar and Arroja, Frederico and Ashdown, Mark and Aumont, J and Baccigalupi, Carlo and Ballardini, M and Banday, Anthony J and Barreiro, RB and Bartolo, Nicola and others},
	journal={Astronomy \& Astrophysics},
	volume={641},
	pages={A1},
	year={2020},
	publisher={EDP sciences}
}

@ARTICLE{KiDS_1000,
	author = {{Asgari}, Marika and {Lin}, Chieh-An and {Joachimi}, Benjamin and {Giblin}, Benjamin and {Heymans}, Catherine and {Hildebrandt}, Hendrik and {Kannawadi}, Arun and {St{\"o}lzner}, Benjamin and {Tr{\"o}ster}, Tilman and {van den Busch}, Jan Luca and {Wright}, Angus H. and {Bilicki}, Maciej and {Blake}, Chris and {de Jong}, Jelte and {Dvornik}, Andrej and {Erben}, Thomas and {Getman}, Fedor and {Hoekstra}, Henk and {K{\"o}hlinger}, Fabian and {Kuijken}, Konrad and {Miller}, Lance and {Radovich}, Mario and {Schneider}, Peter and {Shan}, HuanYuan and {Valentijn}, Edwin},
	title = "{KiDS-1000 cosmology: Cosmic shear constraints and comparison between two point statistics}",
	journal = {\aap},
	year = 2021,
	month = jan,
	volume = {645},
	eid = {A104},
	pages = {A104},
	doi = {10.1051/0004-6361/202039070},
	archivePrefix = {arXiv},
	eprint = {2007.15633},
	primaryClass = {astro-ph.CO},
	adsurl = {https://ui.adsabs.harvard.edu/abs/2021A&A...645A.104A}
}

@article{Brout_2022,
	title={The Pantheon+ Analysis: Cosmological Constraints},
	volume={938},
	ISSN={1538-4357},
	url={http://dx.doi.org/10.3847/1538-4357/ac8e04},
	DOI={10.3847/1538-4357/ac8e04},
	number={2},
	journal={The Astrophysical Journal},
	publisher={American Astronomical Society},
	author={Brout, Dillon and Scolnic, Dan and Popovic, Brodie and Riess, Adam G. and Carr, Anthony and Zuntz, Joe and Kessler, Rick and Davis, Tamara M. and Hinton, Samuel and Jones, David and Kenworthy, W. D’Arcy and Peterson, Erik R. and Said, Khaled and Taylor, Georgie and Ali, Noor and Armstrong, Patrick and Charvu, Pranav and Dwomoh, Arianna and Meldorf, Cole and Palmese, Antonella and Qu, Helen and Rose, Benjamin M. and Sanchez, Bruno and Stubbs, Christopher W. and Vincenzi, Maria and Wood, Charlotte M. and Brown, Peter J. and Chen, Rebecca and Chambers, Ken and Coulter, David A. and Dai, Mi and Dimitriadis, Georgios and Filippenko, Alexei V. and Foley, Ryan J. and Jha, Saurabh W. and Kelsey, Lisa and Kirshner, Robert P. and Möller, Anais and Muir, Jessie and Nadathur, Seshadri and Pan, Yen-Chen and Rest, Armin and Rojas-Bravo, Cesar and Sako, Masao and Siebert, Matthew R. and Smith, Mat and Stahl, Benjamin E. and Wiseman, Phil},
	year={2022},
	month=Oct, pages={110} }

@misc{efstathiou2024,
	title={Challenges to the Lambda CDM Cosmology}, 
	author={George Efstathiou},
	year={2024},
	eprint={2406.12106},
	archivePrefix={arXiv},
	primaryClass={astro-ph.CO},
	url={https://arxiv.org/abs/2406.12106}, 
}

@article{Di_Valentino_2025,
	title={The CosmoVerse White Paper: Addressing observational tensions in cosmology with systematics and fundamental physics},
	volume={49},
	ISSN={2212-6864},
	url={http://dx.doi.org/10.1016/j.dark.2025.101965},
	DOI={10.1016/j.dark.2025.101965},
	journal={Physics of the Dark Universe},
	publisher={Elsevier BV},
	author={Di Valentino and others},
	year={2025},
	month=sep, pages={101965} }

@article{desi2024_dr1_bao,
	title={DESI 2024 VI:  cosmological constraints from the measurements of baryon acoustic oscillations},
	volume={2025},
	ISSN={1475-7516},
	url={http://dx.doi.org/10.1088/1475-7516/2025/02/021},
	DOI={10.1088/1475-7516/2025/02/021},
	number={02},
	journal={Journal of Cosmology and Astroparticle Physics},
	publisher={IOP Publishing},
	author={Adame, A.G. and Aguilar, J. and Ahlen, S. and Alam, S. and Alexander, D.M. and Alvarez, M. and Alves, O. and Anand, A. and Andrade, U. and Armengaud, E. and Avila, S. and Aviles, A. and Awan, H. and Bahr-Kalus, B. and Bailey, S. and Baltay, C. and Bault, A. and Behera, J. and BenZvi, S. and Bera, A. and Beutler, F. and Bianchi, D. and Blake, C. and Blum, R. and Brieden, S. and Brodzeller, A. and Brooks, D. and Buckley-Geer, E. and Burtin, E. and Calderon, R. and Canning, R. and Carnero Rosell, A. and Cereskaite, R. and Cervantes-Cota, J.L. and Chabanier, S. and Chaussidon, E. and Chaves-Montero, J. and Chen, S. and Chen, X. and Claybaugh, T. and Cole, S. and Cuceu, A. and Davis, T.M. and Dawson, K. and de la Macorra, A. and de Mattia, A. and Deiosso, N. and Dey, A. and Dey, B. and Ding, Z. and Doel, P. and Edelstein, J. and Eftekharzadeh, S. and Eisenstein, D.J. and Elliott, A. and Fagrelius, P. and Fanning, K. and Ferraro, S. and Ereza, J. and Findlay, N. and Flaugher, B. and Font-Ribera, A. and Forero-Sánchez, D. and Forero-Romero, J.E. and Frenk, C.S. and Garcia-Quintero, C. and Gaztañaga, E. and Gil-Marín, H. and Gontcho, S.Gontcho A. and Gonzalez-Morales, A.X. and Gonzalez-Perez, V. and Gordon, C. and Green, D. and Gruen, D. and Gsponer, R. and Gutierrez, G. and Guy, J. and Hadzhiyska, B. and Hahn, C. and Hanif, M.M.S. and Herrera-Alcantar, H.K. and Honscheid, K. and Howlett, C. and Huterer, D. and Iršič, V. and Ishak, M. and Juneau, S. and Karaçaylı, N.G. and Kehoe, R. and Kent, S. and Kirkby, D. and Kremin, A. and Krolewski, A. and Lai, Y. and Lan, T.-W. and Landriau, M. and Lang, D. and Lasker, J. and Le Goff, J.M. and Le Guillou, L. and Leauthaud, A. and Levi, M.E. and Li, T.S. and Linder, E. and Lodha, K. and Magneville, C. and Manera, M. and Margala, D. and Martini, P. and Maus, M. and McDonald, P. and Medina-Varela, L. and Meisner, A. and Mena-Fernández, J. and Miquel, R. and Moon, J. and Moore, S. and Moustakas, J. and Mueller, E. and Muñoz-Gutiérrez, A. and Myers, A.D. and Nadathur, S. and Napolitano, L. and Neveux, R. and Newman, J.A. and Nguyen, N.M. and Nie, J. and Niz, G. and Noriega, H.E. and Padmanabhan, N. and Paillas, E. and Palanque-Delabrouille, N. and Pan, J. and Penmetsa, S. and Percival, W.J. and Pieri, M.M. and Pinon, M. and Poppett, C. and Porredon, A. and Prada, F. and Pérez-Fernández, A. and Pérez-Ràfols, I. and Rabinowitz, D. and Raichoor, A. and Ramírez-Pérez, C. and Ramirez-Solano, S. and Rashkovetskyi, M. and Ravoux, C. and Rezaie, M. and Rich, J. and Rocher, A. and Rockosi, C. and Roe, N.A. and Rosado-Marin, A. and Ross, A.J. and Rossi, G. and Ruggeri, R. and Ruhlmann-Kleider, V. and Samushia, L. and Sanchez, E. and Saulder, C. and Schlafly, E.F. and Schlegel, D. and Schubnell, M. and Seo, H. and Shafieloo, A. and Sharples, R. and Silber, J. and Slosar, A. and Smith, A. and Sprayberry, D. and Tan, T. and Tarlé, G. and Taylor, P. and Trusov, S. and Ureña-López, L.A. and Vaisakh, R. and Valcin, D. and Valdes, F. and Vargas-Magaña, M. and Verde, L. and Walther, M. and Wang, B. and Wang, M.S. and Weaver, B.A. and Weaverdyck, N. and Wechsler, R.H. and Weinberg, D.H. and White, M. and Yu, J. and Yu, Y. and Yuan, S. and Yèche, C. and Zaborowski, E.A. and Zarrouk, P. and Zhang, H. and Zhao, C. and Zhao, R. and Zhou, R. and Zhuang, T. and Zou, H. and },
	year={2025},
	month=Feb, pages={021} }

@article{desi2025_dr2_bao,
	title={DESI DR2 results. II. Measurements of baryon acoustic oscillations and cosmological constraints},
	volume={112},
	ISSN={2470-0029},
	url={http://dx.doi.org/10.1103/tr6y-kpc6},
	DOI={10.1103/tr6y-kpc6},
	number={8},
	journal={Physical Review D},
	publisher={American Physical Society (APS)},
	author={Abdul Karim, M. and Aguilar, J. and Ahlen, S. and Alam, S. and Allen, L. and Prieto, C. Allende and Alves, O. and Anand, A. and Andrade, U. and Armengaud, E. and Aviles, A. and Bailey, S. and Baltay, C. and Bansal, P. and Bault, A. and Behera, J. and BenZvi, S. and Bianchi, D. and Blake, C. and Brieden, S. and Brodzeller, A. and Brooks, D. and Buckley-Geer, E. and Burtin, E. and Calderon, R. and Canning, R. and Rosell, A. Carnero and Carrilho, P. and Casas, L. and Castander, F. J. and Charles, M. and Chaussidon, E. and Chaves-Montero, J. and Chebat, D. and Chen, X. and Claybaugh, T. and Cole, S. and Cooper, A. P. and Cuceu, A. and Dawson, K. S. and de la Macorra, A. and de Mattia, A. and Deiosso, N. and Della Costa, J. and Demina, R. and Dey, A. and Dey, B. and Ding, Z. and Doel, P. and Edelstein, J. and Eisenstein, D. J. and Elbers, W. and Fagrelius, P. and Fanning, K. and Fernández-García, E. and Ferraro, S. and Font-Ribera, A. and Forero-Romero, J. E. and Frenk, C. S. and Garcia-Quintero, C. and Garrison, L. H. and Gaztañaga, E. and Gil-Marín, H. and Gontcho, S. Gontcho A. and Gonzalez, D. and Gonzalez-Morales, A. X. and Gordon, C. and Green, D. and Gutierrez, G. and Guy, J. and Hadzhiyska, B. and Hahn, C. and He, S. and Herbold, M. and Herrera-Alcantar, H. K. and Ho, M.-F. and Honscheid, K. and Howlett, C. and Huterer, D. and Ishak, M. and Juneau, S. and Kamble, N. V. and Karaçayl𝚤, N. G. and Kehoe, R. and Kent, S. and Kim, A. G. and Kirkby, D. and Kisner, T. and Koposov, S. E. and Kremin, A. and Krolewski, A. and Lahav, O. and Lamman, C. and Landriau, M. and Lang, D. and Lasker, J. and Le Goff, J. M. and Le Guillou, L. and Leauthaud, A. and Levi, M. E. and Li, Q. and Li, T. S. and Lodha, K. and Lokken, M. and Lozano-Rodríguez, F. and Magneville, C. and Manera, M. and Martini, P. and Matthewson, W. L. and Meisner, A. and Mena-Fernández, J. and Menegas, A. and Mergulhão, T. and Miquel, R. and Moustakas, J. and Muñoz-Gutiérrez, A. and Muñoz-Santos, D. and Myers, A. D. and Nadathur, S. and Naidoo, K. and Napolitano, L. and Newman, J. A. and Niz, G. and Noriega, H. E. and Paillas, E. and Palanque-Delabrouille, N. and Pan, J. and Peacock, J. A. and Ibanez, M. P. and Percival, W. J. and Pérez-Fernández, A. and Pérez-Ràfols, I. and Pieri, M. M. and Poppett, C. and Prada, F. and Rabinowitz, D. and Raichoor, A. and Ramírez-Pérez, C. and Rashkovetskyi, M. and Ravoux, C. and Rich, J. and Rocher, A. and Rockosi, C. and Rohlf, J. and Román-Herrera, J. O. and Ross, A. J. and Rossi, G. and Ruggeri, R. and Ruhlmann-Kleider, V. and Samushia, L. and Sanchez, E. and Sanders, N. and Schlegel, D. and Schubnell, M. and Seo, H. and Shafieloo, A. and Sharples, R. and Silber, J. and Sinigaglia, F. and Sprayberry, D. and Tan, T. and Tarlé, G. and Taylor, P. and Turner, W. and Ureña-López, L. A. and Vaisakh, R. and Valdes, F. and Valogiannis, G. and Vargas-Magaña, M. and Verde, L. and Walther, M. and Weaver, B. A. and Weinberg, D. H. and White, M. and Wolfson, M. and Yèche, C. and Yu, J. and Zaborowski, E. A. and Zarrouk, P. and Zhai, Z. and Zhang, H. and Zhao, C. and Zhao, G. B. and Zhou, R. and Zou, H. and },
	year={2025},
	month=Oct }

@misc{des2026,
	title={Dark Energy Survey Year 6 Results: Cosmological Constraints from Galaxy Clustering and Weak Lensing}, 
	author={{DES Collaboration}},
	year={2026},
	eprint={2601.14559},
	archivePrefix={arXiv},
	primaryClass={astro-ph.CO},
	url={https://arxiv.org/abs/2601.14559}, 
}

@ARTICLE{gravitino_wdm_1997,
	author = {{Kawasaki}, M. and {Sugiyama}, Naoshi and {Yanagida}, T.},
	title = "{Gravitino Warm Dark Matter Motivated by Gauge-Mediated Supersymmetry Breaking Theories}",
	journal = {Modern Physics Letters A},
	year = 1997,
	month = jan,
	volume = {12},
	number = {17},
	pages = {1275-1282},
	doi = {10.1142/S021773239700128X},
	archivePrefix = {arXiv},
	eprint = {hep-ph/9607273},
	primaryClass = {hep-ph},
	adsurl = {https://ui.adsabs.harvard.edu/abs/1997MPLA...12.1275K}
}

@article{Bertone_2005,
	title = {Particle dark matter: evidence, candidates and constraints},
	journal = {Physics Reports},
	volume = {405},
	number = {5},
	pages = {279-390},
	year = {2005},
	issn = {0370-1573},
	doi = {https://doi.org/10.1016/j.physrep.2004.08.031},
	url = {https://www.sciencedirect.com/science/article/pii/S0370157304003515},
	author = {Gianfranco Bertone and Dan Hooper and Joseph Silk}
}

@article{Undagoitia_2015,
	title={Dark matter direct-detection experiments},
	volume={43},
	ISSN={1361-6471},
	url={http://dx.doi.org/10.1088/0954-3899/43/1/013001},
	DOI={10.1088/0954-3899/43/1/013001},
	number={1},
	journal={Journal of Physics G: Nuclear and Particle Physics},
	publisher={IOP Publishing},
	author={Undagoitia, Teresa Marrodán and Rauch, Ludwig},
	year={2015},
	month=Dec, pages={013001} }

@article{Gaskins_2016,
	title={A review of indirect searches for particle dark matter},
	volume={57},
	ISSN={1366-5812},
	url={http://dx.doi.org/10.1080/00107514.2016.1175160},
	DOI={10.1080/00107514.2016.1175160},
	number={4},
	journal={Contemporary Physics},
	publisher={Informa UK Limited},
	author={Gaskins, Jennifer M.},
	year={2016},
	month={June}, pages={496–525} }

@article{Boyarsky_2018,
	author = "Boyarsky, A. and Drewes, M. and Lasserre, T. and Mertens, S. and Ruchayskiy, O.",
	title = "{Sterile neutrino Dark Matter}",
	eprint = "1807.07938",
	archivePrefix = "arXiv",
	primaryClass = "hep-ph",
	doi = "10.1016/j.ppnp.2018.07.004",
	journal = "Prog. Part. Nucl. Phys.",
	volume = "104",
	pages = "1--45",
	year = "2019"
}

@article{Dodelson_1994,
	title = {Sterile neutrinos as dark matter},
	author = {Dodelson, Scott and Widrow, Lawrence M.},
	journal = {Phys. Rev. Lett.},
	volume = {72},
	issue = {1},
	pages = {17--20},
	numpages = {0},
	year = {1994},
	month = {Jan},
	publisher = {American Physical Society},
	doi = {10.1103/PhysRevLett.72.17},
	url = {https://link.aps.org/doi/10.1103/PhysRevLett.72.17}
}

@article{Shi_Fuller_1999,
	title = {New Dark Matter Candidate: Nonthermal Sterile Neutrinos},
	author = {Shi, Xiangdong and Fuller, George M.},
	journal = {Phys. Rev. Lett.},
	volume = {82},
	issue = {14},
	pages = {2832--2835},
	numpages = {0},
	year = {1999},
	month = {Apr},
	publisher = {American Physical Society},
	doi = {10.1103/PhysRevLett.82.2832},
	url = {https://link.aps.org/doi/10.1103/PhysRevLett.82.2832}
}

@article{Pagels_Primack_1982,
	title = {Supersymmetry, Cosmology, and New Physics at Teraelectronvolt Energies},
	author = {Pagels, Heinz and Primack, Joel R.},
	journal = {Phys. Rev. Lett.},
	volume = {48},
	issue = {4},
	pages = {223--226},
	numpages = {0},
	year = {1982},
	month = {Jan},
	publisher = {American Physical Society},
	doi = {10.1103/PhysRevLett.48.223},
	url = {https://link.aps.org/doi/10.1103/PhysRevLett.48.223}
}

@article{Schumann_2019,
	title={Direct detection of WIMP dark matter: concepts and status},
	volume={46},
	ISSN={1361-6471},
	url={http://dx.doi.org/10.1088/1361-6471/ab2ea5},
	DOI={10.1088/1361-6471/ab2ea5},
	number={10},
	journal={Journal of Physics G: Nuclear and Particle Physics},
	publisher={IOP Publishing},
	author={Schumann, Marc},
	year={2019},
	month=Aug, pages={103003} }

@article{Arbey_2021,
	title={Dark matter and the early Universe: A review},
	volume={119},
	ISSN={0146-6410},
	url={http://dx.doi.org/10.1016/j.ppnp.2021.103865},
	DOI={10.1016/j.ppnp.2021.103865},
	journal={Progress in Particle and Nuclear Physics},
	publisher={Elsevier BV},
	author={Arbey, A. and Mahmoudi, F.},
	year={2021},
	month={July}, pages={103865} }

@article{Alam_2001,
 	author = "Alam, S. M. Khairul and Bullock, James S. and Weinberg, David H.",
 	title = "{Dark matter properties and halo central densities}",
 	eprint = "astro-ph/0109392",
 	archivePrefix = "arXiv",
 	doi = "10.1086/340190",
 	journal = "Astrophys. J.",
 	volume = "572",
 	pages = "34--40",
 	year = "2002"
 }

@article{Hogan_2000,
	title = {New dark matter physics: Clues from halo structure},
	author = {Hogan, Craig J. and Dalcanton, Julianne J.},
	journal = {Phys. Rev. D},
	volume = {62},
	issue = {6},
	pages = {063511},
	numpages = {1},
	year = {2000},
	month = {Aug},
	publisher = {American Physical Society},
	doi = {10.1103/PhysRevD.62.063511},
	url = {https://link.aps.org/doi/10.1103/PhysRevD.62.063511}
}

@article{Bode_2001,
	doi = {10.1086/321541},
	url = {https://doi.org/10.1086/321541},
	year = {2001},
	month = {jul},
	publisher = {},
	volume = {556},
	number = {1},
	pages = {93},
	author = {Bode, Paul and Ostriker, Jeremiah P. and Turok, Neil},
	title = {Halo Formation in Warm Dark Matter Models},
	journal = {The Astrophysical Journal}
}

@article{Viel_2005,
	title = {Constraining warm dark matter candidates including sterile neutrinos and light gravitinos with WMAP and the Lyman-$\ensuremath{\alpha}$ forest},
	author = {Viel, Matteo and Lesgourgues, Julien and Haehnelt, Martin G. and Matarrese, Sabino and Riotto, Antonio},
	journal = {Phys. Rev. D},
	volume = {71},
	issue = {6},
	pages = {063534},
	numpages = {10},
	year = {2005},
	month = {Mar},
	publisher = {American Physical Society},
	doi = {10.1103/PhysRevD.71.063534},
	url = {https://link.aps.org/doi/10.1103/PhysRevD.71.063534}
}

@article{Schneider_2012_wdm,
	author = {Schneider, Aurel and Smith, Robert E. and Macciò, Andrea V. and Moore, Ben},
	title = {Non-linear evolution of cosmological structures in warm dark matter models},
	journal = {Monthly Notices of the Royal Astronomical Society},
	volume = {424},
	number = {1},
	pages = {684-698},
	year = {2012},
	month = {07},
	eprint = {https://academic.oup.com/mnras/article-pdf/424/1/684/3296247/mnras0424-0684.pdf},
}

@article{Schneider_2013_wdm,
	author = {Schneider, Aurel and Smith, Robert E. and Reed, Darren},
	title = {Halo mass function and the free streaming scale},
	journal = {Monthly Notices of the Royal Astronomical Society},
	volume = {433},
	number = {2},
	pages = {1573-1587},
	year = {2013},
	month = {06},
	issn = {0035-8711},
	doi = {10.1093/mnras/stt829},
	url = {https://doi.org/10.1093/mnras/stt829},
	eprint = {https://academic.oup.com/mnras/article-pdf/433/2/1573/4926418/stt829.pdf},
}

@article{Ludlow:2016ifl,
	author = "Ludlow, Aaron D. and Bose, Sownak and Angulo, Ra{\'u}l E. and Wang, Lan and Hellwing, Wojciech A. and Navarro, Julio F. and Cole, Shaun and Frenk, Carlos S.",
	title = "{The mass{\textendash}concentration{\textendash}redshift relation of cold and warm dark matter haloes}",
	eprint = "1601.02624",
	archivePrefix = "arXiv",
	primaryClass = "astro-ph.CO",
	doi = "10.1093/mnras/stw1046",
	journal = "Mon. Not. Roy. Astron. Soc.",
	volume = "460",
	number = "2",
	pages = "1214--1232",
	year = "2016"
}

@ARTICLE{Lovell2020a,
	author = {{Lovell}, Mark R.},
	title = "{Toward a General Parameterization of the Warm Dark Matter Halo Mass Function}",
	journal = {\apj},
	year = 2020,
	month = jul,
	volume = {897},
	number = {2},
	eid = {147},
	pages = {147},
	doi = {10.3847/1538-4357/ab982a},
	archivePrefix = {arXiv},
	eprint = {2003.01125},
	primaryClass = {astro-ph.CO},
	adsurl = {https://ui.adsabs.harvard.edu/abs/2020ApJ...897..147L}
}

@article{MiguelRocha_sidm_2013,
	author = {Rocha, Miguel and Peter, Annika H. G. and Bullock, James S. and Kaplinghat, Manoj and Garrison-Kimmel, Shea and Oñorbe, Jose and Moustakas, Leonidas A.},
	title = {Cosmological simulations with self-interacting dark matter – I. Constant-density cores and substructure},
	journal = {Monthly Notices of the Royal Astronomical Society},
	volume = {430},
	number = {1},
	pages = {81-104},
	year = {2013},
	month = {01},
	issn = {0035-8711},
	doi = {10.1093/mnras/sts514},
	url = {https://doi.org/10.1093/mnras/sts514},
	eprint = {https://academic.oup.com/mnras/article-pdf/430/1/81/3064615/sts514.pdf},
}

@article{Hui_2016,
	author = "Hui, Lam and Ostriker, Jeremiah P. and Tremaine, Scott and Witten, Edward",
	title = "{Ultralight scalars as cosmological dark matter}",
	eprint = "1610.08297",
	archivePrefix = "arXiv",
	primaryClass = "astro-ph.CO",
	doi = "10.1103/PhysRevD.95.043541",
	journal = "Phys. Rev. D",
	volume = "95",
	number = "4",
	pages = "043541",
	year = "2017"
}

@article{Tulin_2017,
	author = "Tulin, Sean and Yu, Hai-Bo",
	title = "{Dark Matter Self-interactions and Small Scale Structure}",
	eprint = "1705.02358",
	archivePrefix = "arXiv",
	primaryClass = "hep-ph",
	doi = "10.1016/j.physrep.2017.11.004",
	journal = "Phys. Rept.",
	volume = "730",
	pages = "1--57",
	year = "2018"
}

@article{Banik_2021,
	title={Novel constraints on the particle nature of dark matter from stellar streams},
	volume={2021},
	ISSN={1475-7516},
	url={http://dx.doi.org/10.1088/1475-7516/2021/10/043},
	DOI={10.1088/1475-7516/2021/10/043},
	number={10},
	journal={Journal of Cosmology and Astroparticle Physics},
	publisher={IOP Publishing},
	author={Banik, Nilanjan and Bovy, Jo and Bertone, Gianfranco and Erkal, Denis and de Boer, T.J.L.},
	year={2021},
	month=Oct, pages={043} }

@article{Enzi_2021,
	author = "Enzi, Wolfgang and others",
	title = "{Joint constraints on thermal relic dark matter from strong gravitational lensing, the Ly{\,}{\ensuremath{\alpha}} forest, and Milky Way satellites}",
	eprint = "2010.13802",
	archivePrefix = "arXiv",
	primaryClass = "astro-ph.CO",
	doi = "10.1093/mnras/stab1960",
	journal = "Mon. Not. Roy. Astron. Soc.",
	volume = "506",
	number = "4",
	pages = "5848--5862",
	year = "2021"
}

@article{Nadler_2021,
	title={Dark Matter Constraints from a Unified Analysis of Strong Gravitational Lenses and Milky Way Satellite Galaxies},
	volume={917},
	ISSN={1538-4357},
	url={http://dx.doi.org/10.3847/1538-4357/abf9a3},
	DOI={10.3847/1538-4357/abf9a3},
	number={1},
	journal={The Astrophysical Journal},
	publisher={American Astronomical Society},
	author={Nadler, Ethan O. and Birrer, Simon and Gilman, Daniel and Wechsler, Risa H. and Du, Xiaolong and Benson, Andrew and Nierenberg, Anna M. and Treu, Tommaso},
	year={2021},
	month=Aug, pages={7} }

@article{Chatterjee_2024,
	author = {Chatterjee, Atrideb and Choudhury, Tirthankar Roy},
	title = {Warm Dark matter constraints from the joint analysis of CMB, Ly α, and global 21 cm data},
	journal = {Monthly Notices of the Royal Astronomical Society},
	volume = {527},
	number = {4},
	pages = {10777-10787},
	year = {2024},
	month = {02},
	issn = {0035-8711},
	doi = {10.1093/mnras/stad3930},
	url = {https://doi.org/10.1093/mnras/stad3930},
	eprint = {https://academic.oup.com/mnras/article-pdf/527/4/10777/54945779/stad3930.pdf},
}

@article{Irsic2024,
	title={Unveiling dark matter free streaming at the smallest scales with the high redshift Lyman-alpha forest},
	volume={109},
	ISSN={2470-0029},
	url={http://dx.doi.org/10.1103/PhysRevD.109.043511},
	DOI={10.1103/physrevd.109.043511},
	number={4},
	journal={Physical Review D},
	publisher={American Physical Society (APS)},
	author={Iršič, Vid and Viel, Matteo and Haehnelt, Martin G. and Bolton, James S. and Molaro, Margherita and Puchwein, Ewald and Boera, Elisa and Becker, George D. and Gaikwad, Prakash and Keating, Laura C. and Kulkarni, Girish},
	year={2024},
	month=Feb}

@article{Liu_2024,
	doi = {10.3847/1538-4357/ad4ed8},
	url = {https://doi.org/10.3847/1538-4357/ad4ed8},
	year = {2024},
	month = {jun},
	publisher = {The American Astronomical Society},
	volume = {968},
	number = {2},
	pages = {79},
	author = {Liu, Bin and Shan, Huanyuan and Zhang, Jiajun},
	title = {New Galaxy UV Luminosity Constraints on Warm Dark Matter from JWST},
	journal = {The Astrophysical Journal}
}

@misc{gilman2026jwstlensedquasardark,
	title={JWST lensed quasar dark matter survey IV: Stringent warm dark matter constraints from the joint reconstruction of extended lensed arcs and quasar flux ratios}, 
	author={D. Gilman and A. M. Nierenberg and T. Treu and C. Gannon and X. Du and H. Paugnat and S. Birrer and A. J. Benson and P. Mozumdar and K. C. Wong and D. Williams and R. E. Keeley and K. N. Abazajian and T. Anguita and V. N. Bennert and S. G. Djorgovski and S. H. Hoenig and A. Kusenko and M. Malkan and T. Morishita and V. Motta and L. A. Moustakas and W. Sheu and D. Sluse and D. Stern and M. Stiavelli},
	year={2026},
	eprint={2511.07513},
	archivePrefix={arXiv},
	primaryClass={astro-ph.CO},
	url={https://arxiv.org/abs/2511.07513}, 
}

@misc{liu2_2026,
	title={Joint Constraints on Fuzzy and Warm Dark Matter from Satellite Populations of the Milky Way and Andromeda}, 
	author={Jianxiang Liu and Yan Gong and Kai Liao},
	year={2026},
	eprint={2512.01361},
	archivePrefix={arXiv},
	primaryClass={astro-ph.CO},
	url={https://arxiv.org/abs/2512.01361}, 
}

@techreport{LIGO-M1100296-v2,
    author      = {Iyer, Bala and Souradeep, Tarun and Unnikrishnan, C. S. and Dhurandhar, Sanjeev and Raja, Sendhil and Sengupta, Anand},
    title       = {{LIGO-India, Proposal of the Consortium for Indian Initiative in Gravitational-wave Observations (IndIGO)}},
    institution = {LIGO Scientific Collaboration},
    number      = {LIGO-M1100296-v2},
    year        = {2011},
    month       = {Nov},
    url         = {https://dcc-llo.ligo.org/LIGO-M1100296/public}
}

@article{hild2011sensitivity,
	title={Sensitivity studies for third-generation gravitational wave observatories},
	author={Hild, S and Abernathy, M and Acernese, F ea and Amaro-Seoane, P and Andersson, N and Arun, K and Barone, F and Barr, B and Barsuglia, M and Beker, M and others},
	journal={Classical and Quantum gravity},
	volume={28},
	number={9},
	pages={094013},
	year={2011},
	publisher={IOP Publishing}
}

@article{acernese2014advanced,
	title={Advanced Virgo: a second-generation interferometric gravitational wave detector},
	author={Acernese, Fausto and Agathos, M and Agatsuma, K and Aisa, Damiano and Allemandou, N and Allocca, Aea and Amarni, J and Astone, Pia and Balestri, G and Ballardin, G and others},
	journal={Classical and Quantum Gravity},
	volume={32},
	number={2},
	pages={024001},
	year={2014},
	publisher={IOP Publishing}
}

@article{aasi2015advanced,
	title={Advanced ligo},
	author={Aasi, Junaid and Abbott, BP and Abbott, Richard and Abbott, Thomas and Abernathy, MR and Ackley, Kendall and Adams, Carl and Adams, Thomas and Addesso, Paolo and Adhikari, RX and others},
	journal={Classical and quantum gravity},
	volume={32},
	number={7},
	pages={074001},
	year={2015},
	publisher={IOP Publishing}
}

@misc{Reitze_2019,
	title = {Cosmic Explorer: The U.S. Contribution to Gravitational-Wave Astronomy beyond LIGO},
	author = "Reitze, David and others",
	eprint = "1907.04833",
	archivePrefix = "arXiv",
	primaryClass = "astro-ph.IM",
	reportNumber = "LIGO-P1900316",
	journal = "Bull. Am. Astron. Soc.",
	volume = "51",
	number = "7",
	pages = "035",
	year = "2019"
}

@misc{H1L1V1-psd-O3O4O5,
	author = "{LIGO-Virgo Collaboration}",
	title = "Noise curves used for Simulations in the update of the Observing Scenarios Paper",
	note = "{LIGO} Document T2000012-v1",
	url = "https://dcc.ligo.org/LIGO-T2000012/public",
	year ="2020"
}

@article{maggiore2020science,
	title={Science case for the Einstein telescope},
	author={Maggiore, Michele and Van Den Broeck, Chris and Bartolo, Nicola and Belgacem, Enis and Bertacca, Daniele and Bizouard, Marie Anne and Branchesi, Marica and Clesse, Sebastien and Foffa, Stefano and Garc{\'\i}a-Bellido, Juan and others},
	journal={Journal of Cosmology and Astroparticle Physics},
	volume={2020},
	number={03},
	pages={050},
	year={2020},
	publisher={IOP Publishing}
}

@article{Kalogera:2021bya,
	author = "Kalogera, Vicky and others",
	title = "{The Next Generation Global Gravitational Wave Observatory: The Science Book}",
	eprint = "2111.06990",
	archivePrefix = "arXiv",
	primaryClass = "gr-qc",
	month = "11",
	year = "2021"
}

@article{Saleem_2022,
	doi = {10.1088/1361-6382/ac3b99},
	url = {https://doi.org/10.1088/1361-6382/ac3b99},
	year = {2021},
	month = {dec},
	publisher = {IOP Publishing},
	volume = {39},
	number = {2},
	pages = {025004},
	author = {Saleem, M and Rana, Javed and Gayathri, V and Vijaykumar, Aditya and Goyal, Srashti and Sachdev, Surabhi and Suresh, Jishnu and Sudhagar, S and Mukherjee, Arunava and Gaur, Gurudatt and Sathyaprakash, Bangalore and Pai, Archana and Adhikari, Rana X and Ajith, P and Bose, Sukanta},
	title = {The science case for LIGO-India},
	journal = {Classical and Quantum Gravity}
}

@misc{evans2021horizonstudycosmicexplorer,
	title={A Horizon Study for Cosmic Explorer: Science, Observatories, and Community}, 
	author={Matthew Evans and others},
	year={2021},
	eprint={2109.09882},
	archivePrefix={arXiv},
	primaryClass={astro-ph.IM},
	url={https://arxiv.org/abs/2109.09882}, 
}

@misc{HLA-psd-O6,
	author       = {{LIGO-Virgo Collaboration}},
	title        = {A{\#} {S}train {S}ensitivity},
	year         = {2023},
	note         = "{LIGO} Document T2300041-v1",
	url          = {https://dcc.ligo.org/LIGO-T2300041/public},
	urldate      = {2025-08-28},
}

@misc{HLA-voyager,
	author       = {{LIGO-Virgo Collaboration}},
	title        = {Report from the LSC Post-O5 Study Group},
	year         = {2023},
	note         = {{LIGO} Document T2200287–v2},
	url          = {https://dcc.ligo.org/public/0183/T2200287/002/T2200287v2_PO5report.pdf},
}

@misc{HLV-psd-O4a,
	author       = {{LIGO-Virgo Collaboration}},
	title        = {GWIStat},
	year       = {2023},
	url          = {https://git.ligo.org/computing/services/gwistat/-/tree/master/psd},
	urldate      = {2025-08-28},
}

@misc{ligocurrentPlan,
	author       = {{LIGO Scientific Collaboration}},
	title        = {Observing Plans},
	year         = {2026},
	howpublished = {\url{https://observing.docs.ligo.org/plan/}},
	note         = {Accessed: 17 July 2026}
}

@article{abbott2019gwtc,
	title={GWTC-1: a gravitational-wave transient catalog of compact binary mergers observed by LIGO and Virgo during the first and second observing runs},
	author={Abbott, Benjamin P and Abbott, Richard and Abbott, TDea and Abraham, S and Acernese, F and Ackley, K and Adams, C and Adhikari, RX and Adya, VB and Affeldt, Christoph and others},
	journal={Physical Review X},
	volume={9},
	number={3},
	pages={031040},
	year={2019},
	publisher={APS}
}

@article{nitz20191,
	title={1-OGC: The first open gravitational-wave catalog of binary mergers from analysis of public Advanced LIGO data},
	author={Nitz, Alexander H and Capano, Collin and Nielsen, Alex B and Reyes, Steven and White, Rebecca and Brown, Duncan A and Krishnan, Badri},
	journal={The Astrophysical Journal},
	volume={872},
	number={2},
	pages={195},
	year={2019},
	publisher={IOP Publishing}
}

@article{nitz20202,
	title={2-OGC: Open Gravitational-wave Catalog of binary mergers from analysis of public Advanced LIGO and Virgo data},
	author={Nitz, Alexander H and Dent, Thomas and Davies, Gareth S and Kumar, Sumit and Capano, Collin D and Harry, Ian and Mozzon, Simone and Nuttall, Laura and Lundgren, Andrew and T{\'a}pai, M{\'a}rton},
	journal={The Astrophysical Journal},
	volume={891},
	number={2},
	pages={123},
	year={2020},
	publisher={American Astronomical Society}
}

@article{venumadhav2020new,
	title={New binary black hole mergers in the second observing run of Advanced LIGO and Advanced Virgo},
	author={Venumadhav, Tejaswi and Zackay, Barak and Roulet, Javier and Dai, Liang and Zaldarriaga, Matias},
	journal={Physical Review D},
	volume={101},
	number={8},
	pages={083030},
	year={2020},
	publisher={APS}
}

@article{abbott2021gwtc,
	title={GWTC-2: compact binary coalescences observed by LIGO and Virgo during the first half of the third observing run},
	author={Abbott, Richard and Abbott, TD and Abraham, S and Acernese, F and Ackley, K and Adams, A and Adams, C and Adhikari, RX and Adya, VB and Affeldt, Christoph and others},
	journal={Physical Review X},
	volume={11},
	number={2},
	pages={021053},
	year={2021},
	publisher={APS}
}

@article{nitz20213,
	title={3-OGC: Catalog of gravitational waves from compact-binary mergers},
	author={Nitz, Alexander H and Capano, Collin D and Kumar, Sumit and Wang, Yi-Fan and Kastha, Shilpa and Sch{\"a}fer, Marlin and Dhurkunde, Rahul and Cabero, Miriam},
	journal={The Astrophysical Journal},
	volume={922},
	number={1},
	pages={76},
	year={2021},
	publisher={IOP Publishing}
}

@article{zackay2021detecting,
	title={Detecting gravitational waves with disparate detector responses: Two new binary black hole mergers},
	author={Zackay, Barak and Dai, Liang and Venumadhav, Tejaswi and Roulet, Javier and Zaldarriaga, Matias},
	journal={Physical Review D},
	volume={104},
	number={6},
	pages={063030},
	year={2021},
	publisher={APS}
}

@article{olsen2022new,
	title={New binary black hole mergers in the LIGO-Virgo O3a data},
	author={Olsen, Seth and Venumadhav, Tejaswi and Mushkin, Jonathan and Roulet, Javier and Zackay, Barak and Zaldarriaga, Matias},
	journal={Physical Review D},
	volume={106},
	number={4},
	pages={043009},
	year={2022},
	publisher={APS}
}

@article{ligo2023gwtc,
   title={GWTC-3: Compact Binary Coalescences Observed by LIGO and Virgo during the Second Part of the Third Observing Run},
   volume={13},
   ISSN={2160-3308},
   url={http://dx.doi.org/10.1103/PhysRevX.13.041039},
   DOI={10.1103/physrevx.13.041039},
   number={4},
   journal={Physical Review X},
   publisher={American Physical Society (APS)},
   author={Abbott, R. and Abbott, T. D. and others},
   year={2023},
   month=Dec }

@article{nitz20234,
	title={4-OGC: Catalog of gravitational waves from compact binary mergers},
	author={Nitz, Alexander H and Kumar, Sumit and Wang, Yi-Fan and Kastha, Shilpa and Wu, Shichao and Sch{\"a}fer, Marlin and Dhurkunde, Rahul and Capano, Collin D},
	journal={The Astrophysical Journal},
	volume={946},
	number={2},
	pages={59},
	year={2023},
	publisher={IOP Publishing}
}

@article{wadekar2023new,
	title={New black hole mergers in the LIGO-Virgo O3 data from a gravitational wave search including higher-order harmonics},
	author={Wadekar, Digvijay and Roulet, Javier and Venumadhav, Tejaswi and Mehta, Ajit Kumar and Zackay, Barak and Mushkin, Jonathan and Olsen, Seth and Zaldarriaga, Matias},
	journal={arXiv preprint arXiv:2312.06631},
	year={2023}
}

@article{abbott2024gwtc,
	title={GWTC-2.1: Deep extended catalog of compact binary coalescences observed by LIGO and Virgo during the first half of the third observing run},
	author={Abbott, R and Abbott, TD and Acernese, F and Ackley, K and Adams, C and Adhikari, N and Adhikari, RX and Adya, VB and Affeldt, C and Agarwal, D and others},
	journal={Physical Review D},
	volume={109},
	number={2},
	pages={022001},
	year={2024},
	publisher={APS}
}

@article{abac2025gwtc,
	title={GWTC-4.0: An Introduction to Version 4.0 of the Gravitational-Wave Transient Catalog},
	author={Abac, AG and Abouelfettouh, I and Acernese, F and Ackley, K and Adhicary, S and Adhikari, D and Adhikari, N and Adhikari, RX and Adkins, VK and Afroz, S and others},
	journal={arXiv preprint arXiv:2508.18080},
	year={2025}
}

@article{koloniari2025new,
	title={New gravitational wave discoveries enabled by machine learning},
	author={Koloniari, Alexandra E and Koursoumpa, Evdokia C and Nousi, Paraskevi and Lampropoulos, Paraskevas and Passalis, Nikolaos and Tefas, Anastasios and Stergioulas, Nikolaos},
	journal={Machine Learning: Science and Technology},
	volume={6},
	number={1},
	pages={015054},
	year={2025},
	publisher={IOP Publishing}
}

@article{mehta2025new,
	title={New binary black hole mergers in the LIGO-Virgo O3b data},
	author={Mehta, Ajit Kumar and Olsen, Seth and Wadekar, Digvijay and Roulet, Javier and Venumadhav, Tejaswi and Mushkin, Jonathan and Zackay, Barak and Zaldarriaga, Matias},
	journal={Physical Review D},
	volume={111},
	number={2},
	pages={024049},
	year={2025},
	publisher={APS}
}

@misc{abac2026gwtc,
	title={GWTC-5.0: An Introduction to Version 5.0 of the Gravitational-Wave Transient Catalog}, 
	author={A. G. Abac and others},
	year={2026},
	eprint={2605.27223},
	archivePrefix={arXiv},
	primaryClass={gr-qc},
	url={https://arxiv.org/abs/2605.27223}, 
}

@article{Chen:2024gdn,
    author = "Chen, Hsin-Yu and Ezquiaga, Jose Mar{\'\i}a and Gupta, Ish",
    title = "{Cosmography with next-generation gravitational wave detectors}",
    eprint = "2402.03120",
    archivePrefix = "arXiv",
    primaryClass = "gr-qc",
    doi = "10.1088/1361-6382/ad424f",
    journal = "Class. Quant. Grav.",
    volume = "41",
    number = "12",
    pages = "125004",
    year = "2024"
}

@article{hannuksela2019search,
	title        = {Search for gravitational lensing signatures in LIGO-Virgo binary black hole events},
	author       = {Hannuksela, O.A. and Haris, K. and Ng, K.K.Y. and Kumar, S. and Mehta, A.K. and Keitel, D. and Li, T.G.F. and Ajith, P.},
	year         = 2019,
	journal      = {Astrophys. J. Lett.},
	volume       = 874,
	number       = 1,
	pages        = {L2},
	doi          = {10.3847/2041-8213/ab0c0f},
	archiveprefix = {arXiv},
	eprint       = {1901.02674},
	primaryclass = {gr-qc},
	reportnumber = {LIGO Document P1800297, LIGO-P1800297}
}

@article{dai2020search,
	title        = {Search for Lensed Gravitational Waves Including Morse Phase Information: An Intriguing Candidate in O2},
	author       = {Dai, Liang and Zackay, Barak and Venumadhav, Tejaswi and Roulet, Javier and Zaldarriaga, Matias},
	year         = 2020,
	month        = 7,
	journal={arXiv preprint arXiv:2007.12709},
	archiveprefix = {arXiv},
	eprint       = {2007.12709},
	primaryclass = {astro-ph.HE}
}

@article{mcisaac2020search,
	title        = {{Search for strongly lensed counterpart images of binary black hole mergers in the first two LIGO observing runs}},
	author       = {McIsaac, Connor and Keitel, David and Collett, Thomas and Harry, Ian and Mozzon, Simone and Edy, Oliver and Bacon, David},
	year         = 2020,
	journal      = {Phys. Rev. D},
	volume       = 102,
	number       = 8,
	pages        = {084031},
	doi          = {10.1103/PhysRevD.102.084031},
	eprint       = {1912.05389},
	archiveprefix = {arXiv},
	primaryclass = {gr-qc},
	reportnumber = {LIGO-P1900360}
}

@article{LIGOScientific:2021izm,
	title        = {Search for Lensing Signatures in the Gravitational-Wave Observations from the First Half of LIGO\textendash{}Virgo\textquoteright{}s Third Observing Run},
	author       = {Abbott, R. and others},
	year         = 2021,
	month        = 5,
	journal      = {Astrophys. J.},
	volume       = 923,
	number       = 1,
	pages        = 14,
	doi          = {10.3847/1538-4357/ac23db},
	collaboration = {LIGO Scientific, VIRGO},
	eprint       = {2105.06384},
	archiveprefix = {arXiv},
	primaryclass = {gr-qc},
	reportnumber = {LIGO-P2000400}
}

@article{janquart2023follow,
	title={Follow-up analyses to the O3 LIGO--Virgo--KAGRA lensing searches},
	volume={526},
	ISSN={1365-2966},
	url={http://dx.doi.org/10.1093/mnras/stad2909},
	DOI={10.1093/mnras/stad2909},
	number={3},
	journal={Monthly Notices of the Royal Astronomical Society},
	publisher={Oxford University Press (OUP)},
	author={Janquart, J and others},
	year={2023},
	month=sep,
	pages={3832–3860}
}

@article{li2023targeted,
	author = "Li, Alvin K. Y. and Lo, Rico K. L. and Sachdev, Surabhi and Chan, Juno C. L. and Lin, E. T. and Li, Tjonnie G. F. and Weinstein, Alan J.",
	collaboration = "LIGO Scientific, Virgo",
	title = "{Targeted subthreshold search for strongly lensed gravitational-wave events}",
	eprint = "1904.06020",
	archivePrefix = "arXiv",
	primaryClass = "gr-qc",
	doi = "10.1103/PhysRevD.107.123014",
	journal = "Phys. Rev. D",
	volume = "107",
	number = "12",
	pages = "123014",
	year = "2023"
}

@article{abbott2023search,
	doi = {10.3847/1538-4357/ad3e83},
	url = {https://dx.doi.org/10.3847/1538-4357/ad3e83},
	year = {2024},
	month = {jul},
	publisher = {The American Astronomical Society},
	volume = {970},
	number = {2},
	pages = {191},
	author = {R. Abbott and H. Abe and F. Acernese and K. Ackley and others},
	title = {Search for Gravitational-lensing Signatures in the Full Third Observing Run of the LIGO–Virgo Network},
	journal = {The Astrophysical Journal}
}

@misc{ligo_scientific_collaboration_and_virgo_2024_10841987,
	author       = {Abbott, R and Abe, H and Acernese, F and Ackley, K and Adhicary, S and Adhikari, N and Adhikari, RX and Adkins, VK and Adya, VB and Affeldt, C and others},
	title        = {{The data for "Search for gravitational-lensing
	signatures in the full third observing run of the
	LIGO–Virgo network"}},
	month        = mar,
	year         = 2024,
	publisher    = {Zenodo},
	doi          = {10.5281/zenodo.10841987},
	url          = {https://doi.org/10.5281/zenodo.10841987}
}

@article{abac2025gwtclens,
	title={GWTC-4.0: Searches for Gravitational-Wave Lensing Signatures},
	author={Abac, AG and Abouelfettouh, I and Acernese, F and Ackley, K and Adamcewicz, C and Adhicary, S and Adhikari, D and Adhikari, N and Adhikari, RX and Adkins, VK and others},
	journal={arXiv preprint arXiv:2512.16347},
	year={2025}
}

@ARTICLE{Treu_Koopmans2004,
	author = {{Treu}, Tommaso and {Koopmans}, L{\'e}on V.~E.},
	title = "{Massive Dark Matter Halos and Evolution of Early-Type Galaxies to z \raisebox{-0.5ex}\textasciitilde 1}",
	journal = {\apj},
	year = 2004,
	month = aug,
	volume = {611},
	number = {2},
	pages = {739-760},
	doi = {10.1086/422245},
	archivePrefix = {arXiv},
	eprint = {astro-ph/0401373},
	primaryClass = {astro-ph},
	adsurl = {https://ui.adsabs.harvard.edu/abs/2004ApJ...611..739T}
}

@article{Koopmans_2006,
	doi = {10.1086/505696},
	url = {https://doi.org/10.1086/505696},
	year = {2006},
	month = {oct},
	publisher = {},
	volume = {649},
	number = {2},
	pages = {599},
	author = {Koopmans, Léon V. E. and Treu, Tommaso and Bolton, Adam S. and Burles, Scott and Moustakas, Leonidas A.},
	title = {The Sloan Lens ACS Survey. III. The Structure and Formation of Early-Type Galaxies and Their Evolution since z ≈ 1},
	journal = {The Astrophysical Journal}
}

@ARTICLE{Koopmans_2009,
	author = {{Koopmans}, L.~V.~E. and {Bolton}, A. and {Treu}, T. and {Czoske}, O. and {Auger}, M.~W. and {Barnab{\`e}}, M. and {Vegetti}, S. and {Gavazzi}, R. and {Moustakas}, L.~A. and {Burles}, S.},
	title = "{The Structure and Dynamics of Massive Early-Type Galaxies: On Homology, Isothermality, and Isotropy Inside One Effective Radius}",
	journal = {\apjl},
	year = 2009,
	month = sep,
	volume = {703},
	number = {1},
	pages = {L51-L54},
	doi = {10.1088/0004-637X/703/1/L51},
	archivePrefix = {arXiv},
	eprint = {0906.1349},
	primaryClass = {astro-ph.CO},
	adsurl = {https://ui.adsabs.harvard.edu/abs/2009ApJ...703L..51K}
}

@misc{Behroozi_2013,
	doi = {10.1088/0004-637X/770/1/57},
	url = {https://dx.doi.org/10.1088/0004-637X/770/1/57},
	year = {2013},
	month = {may},
	publisher = {The American Astronomical Society},
	volume = {770},
	number = {1},
	pages = {57},
	author = {Behroozi, Peter S. and Wechsler, Risa H. and Conroy, Charlie},
	title = {THE AVERAGE STAR FORMATION HISTORIES OF GALAXIES IN DARK MATTER HALOS FROM z = 0–8},
	journal = {The Astrophysical Journal}
}

@article{Xu_2022,
	doi = {10.3847/1538-4357/ac58f8},
	url = {https://doi.org/10.3847/1538-4357/ac58f8},
	year = {2022},
	month = {apr},
	publisher = {The American Astronomical Society},
	volume = {929},
	number = {1},
	pages = {9},
	author = {Xu, Fei and Ezquiaga, Jose María and Holz, Daniel E.},
	title = {Please Repeat: Strong Lensing of Gravitational Waves as a Probe of Compact Binary and Galaxy Populations},
	journal = {The Astrophysical Journal}
}

@article{Etherington_2023,
	author = {Etherington, Amy and Nightingale, James W and Massey, Richard and Robertson, Andrew and Cao, XiaoYue and Amvrosiadis, Aristeidis and Cole, Shaun and Frenk, Carlos S and He, Qiuhan and Lagattuta, David J and Lange, Samuel and Li, Ran},
	title = {Beyond the bulge–halo conspiracy? Density profiles of early-type galaxies from extended-source strong lensing},
	journal = {Monthly Notices of the Royal Astronomical Society},
	volume = {521},
	number = {4},
	pages = {6005-6018},
	year = {2023},
	month = {06},
	issn = {0035-8711},
	doi = {10.1093/mnras/stad582},
	url = {https://doi.org/10.1093/mnras/stad582},
	eprint = {https://academic.oup.com/mnras/article-pdf/521/4/6005/49845933/stad582.pdf},
}

@ARTICLE{Abe2025,
	author = {{Abe}, Katsuya T. and {Oguri}, Masamune and {Birrer}, Simon and {Khadka}, Narayan and {Marshall}, Philip J. and {Lemon}, Cameron and {More}, Anupreeta and {LSST Dark Energy Science Collaboration}},
	title = "{A halo model approach for mock catalogs of time-variable strong gravitational lenses}",
	journal = {The Open Journal of Astrophysics},
	year = 2025,
	month = jan,
	volume = {8},
	eid = {8},
	pages = {8},
	doi = {10.33232/001c.128482},
	archivePrefix = {arXiv},
	eprint = {2411.07509},
	primaryClass = {astro-ph.CO},
	adsurl = {https://ui.adsabs.harvard.edu/abs/2025OJAp....8E...8A}
}

@misc{Dominik_2013,
	title={DOUBLE COMPACT OBJECTS. II. COSMOLOGICAL MERGER RATES},
	volume={779},
	ISSN={1538-4357},
	url={http://dx.doi.org/10.1088/0004-637X/779/1/72},
	DOI={10.1088/0004-637x/779/1/72},
	number={1},
	journal={The Astrophysical Journal},
	publisher={American Astronomical Society},
	author={Dominik, Michal and Belczynski, Krzysztof and Fryer, Christopher and Holz, Daniel E. and Berti, Emanuele and Bulik, Tomasz and Mandel, Ilya and O’Shaughnessy, Richard},
	year={2013},
	month=nov, pages={72} 
}

@misc{Madau_2014,
	title={Cosmic Star-Formation History},
	volume={52},
	ISSN={1545-4282},
	url={http://dx.doi.org/10.1146/annurev-astro-081811-125615},
	DOI={10.1146/annurev-astro-081811-125615},
	number={1},
	journal={Annual Review of Astronomy and Astrophysics},
	publisher={Annual Reviews},
	author={Madau, Piero and Dickinson, Mark},
	year={2014},
	month=aug, pages={415–486} }

@book{1992schneider,
	author = {{Schneider}, Peter and {Ehlers}, J{\"u}rgen and {Falco}, Emilio E.},
	title = {Gravitational Lenses},
	year = {1992},
	doi = {10.1007/978-3-662-03758-4},
	adsurl = {https://ui.adsabs.harvard.edu/abs/1992grle.book.....S},
	publisher = {Springer Berlin, Heidelberg}
}

@article{Thain:2005,
	author = {Douglas Thain and Todd Tannenbaum and Miron Livny},
	title = {Distributed computing in practice: {The Condor} experience},
	journal = {Concurrency and Computation: Practice and Experience},
	volume = {17},
	number = {2-4},
	pages = {323--356},
	year = {2005},
	publisher = {John Wiley & Sons, Ltd.}
}

@Article{Hunter:2007,
	Author    = {Hunter, J. D.},
	Title     = {Matplotlib: A 2D graphics environment},
	Journal   = {Computing in Science \& Engineering},
	Volume    = {9},
	Number    = {3},
	Pages     = {90--95},
	publisher = {IEEE COMPUTER SOC},
	doi       = {10.1109/MCSE.2007.55},
	year      = 2007
}

@article{barsode2026search,
	title={Search for strong lensing of gravitational waves in the binary black hole events from O1-O4a},
	author={Barsode, Ankur and Maity, Koustav N. and Ajith, P.},
	journal={arXiv preprint arXiv:2607.08466},
	year={2026}
}

\end{document}